\documentclass[11pt,tightenlines,eqsecnum,floats,aps,amsmath,amssymb,nofootinbib,prd,shownopacs,notitlepage]{revtex4}

\usepackage{ragged2e}
\usepackage{graphicx}
\usepackage{soul}
\usepackage{lscape}
\usepackage[utf8]{inputenc}
\usepackage{amsmath,amsfonts,amssymb}
\usepackage{color}
\usepackage[hidelinks]{hyperref}
\hypersetup{
  colorlinks   = true, 
  urlcolor     = blue, 
  linkcolor    = blue, 
  citecolor   = blue 
}
\usepackage{multirow}

\def\newvchanges{\textcolor{black}}

\begin{document}

\title{Power Suppression and Lensing Anomaly -- A phenomenological investigation}
\author{Roshna K}
\email{roshnak.217ph005@nitk.edu.in }
\author{V. Sreenath}
\email{sreenath@nitk.edu.in}
\affiliation{Department of Physics, National Institute of Technology Karnataka, Surathkal, Mangaluru 575025, India.}


\begin{abstract}
Primordial power spectra with low power at long wavelengths can alleviate lensing anomaly. However the extent to which data favours such a primordial spectra is not clear. In this work, we investigate power suppression and related mitigation of lensing anomaly with the help of phenomenological models which are valid over scales of interest.
We consider simple extensions to nearly scale invariant power spectra such as those which includes running and running of running of spectral index. 
We perform Bayesian analysis of these models, which are agnostic about power suppression, 
with Planck legacy data and show that data tend to choose parameters which leads to power suppression at low multipoles. 
We then investigate the connection between power suppression and alleviation of lensing anomaly and show that lensing anomaly is mitigated the most in models with maximum suppression of power at low multipoles. 
We also analyse the significance of these findings using information criteria. These results are further analyzed in the light of Planck Release 4 data using \texttt{CamSpec}, \texttt{HiLLiPoP} and \texttt{LoLLiPoP} likelihoods in which departure of lensing parameter from one is significantly reduced. 
Furthermore, we investigate the ability of near-ultimate future CMB missions such as ECHO to put tighter constraints on these models and to settle the issue. We conclude that we can make stronger conclusions about 
the presence of power suppression in the future by studying such simple phenomenological models.
\end{abstract}

\maketitle
\section{Introduction}\label{sec:1}
Some of the most intriguing questions in cosmology concerns the {\it origin} of our Universe. For instance, we are yet to understand the exact mechanism that generated primordial perturbations that lead to the observed anisotropies and inhomogeneities in the Universe. Though inflationary scenario provides a very simple and elegant solution to this question, it is far from settled. Questions concerning model of inflation, duration of inflation, tensor modes etc. are unresolved \cite{Planck:2018jri, Martin:2013nzq, Martin:2013tda, Steinhardetal, ResponsetoSteinhardetal}. Further, we do not know much about our Universe close to the Planck regime. 
\par 
A way to gain knowledge about the earliest epochs of our Universe is to study the imprints left by primordial perturbations generated then in cosmic microwave background (CMB).
Since our Universe is expanding, physical wavelength of primordial perturbations which are generated early on in the universe expands more and leave their imprints at longer length scales. Hence, the largest angular scales or the lowest multipoles of the CMB contain valuable information about the early universe. 
Ever since the serendipitous discovery of CMB by Penzias and Wilson in 1964 \cite{Penzias:1965wn}, there have been several efforts to study CMB. 
The lowest multipoles of the CMB was first observed by COBE \cite{Bennett:1996ce} satellite. These measurements were then refined considerably by both ground based and satellite missions. For a history of CMB measurements, see, for instance, \cite{Peebles:2009zz}. The current state-of-the-art measurements of anisotropies, especially at low multipoles, in CMB has come from the  Planck mission \cite{Planck:2018nkj}. 
\par 
Bayesian studies of angular power spectra of CMB temperature and polarisation measured by Planck are in fairly good agreement with the predictions of standard model of cosmology especially so at high multipoles \cite{Planck:2018jri, Planck:2018vyg}. 
However, there are certain signals at large angular scales in the CMB which depart mildly, at about $2$ to $3-\sigma$, from the predictions of statistical isotropy and homogeneity by the standard model of cosmology.
These mild departures from the predictions of standard model of cosmology are referred to as CMB anomalies \cite{WMAP:2003ivt, Schwarz:2015cma, Planck:2019evm}. 
They include lack of power at large angular scales, dipolar modulation of low multipoles, alignment of low multipoles, preference for odd parity etc. 
Of these a lack of power at low multipoles has been observed ever since the first measurement of temperature power spectrum by COBE \cite{Hinshaw:1996ut}. 
These features are real, since they have been observed across different satellite missions and hence are unlikely to be due to unmitigated instrumental noise. Further, even though individually they are mild departures from predictions of standard model, their combined occurrence makes them interesting. Even so, we can reason them away by considering our Universe to be a rather rare realization of the Universe predicted by standard model. However, we can be more {\it optimistic} and consider these anomalies to be signals of physics beyond standard model. Since physics of the early universe leave their imprints at low multipoles, there is a possibility that anomalies can be viewed as potential smoking guns of physics of early universe. 
\par 
There have been several studies explaining ways a single CMB anomaly can arise from a primordial scenario. For instance, there have been various works which investigate models which lead to suppression of power at large angular scales \cite{ Vilenkin:1982wt, Contaldi:2003zv, Starobinsky:1992ts, Jain:2008dw, Ragavendra:2020old} (see also \cite{Hazra:2012yn, Sreenath:2013xra, Sreenath:2014nca, Sreenath:2014nka}), dipolar modulation \cite{Prunet:2004zy}, alignment of low multiples \cite{Gordon:2005ai}, quadrupolar modulation \cite{Agullo:2020iqv, Agullo:2020wur} etc. If a single primordial mechanism can explain more than one anomalous feature in the CMB then the possibility that CMB anomalies are signals of primordial physics beyond the standard model becomes more credible. 
Recently, it was shown in the context of loop quantum cosmology that quantum effects could lead to a suppression of power at large angular scales which in turn could lead to lensing anomaly \cite{ Ashtekar:2020gec, Ashtekar:2021izi, Martin-Benito:2023nky}. It was also proposed that non-Gaussian modulation could explain more than one anomaly including dipolar modulation, power suppression and lensing anomaly \cite{Agullo:2021oqk, Agullo:2020cvg, Agullo:2020fbw}. In this context see also, \cite{Delgado:2021mxu, vanTent:2022vgy, Agullo:2017eyh, Sreenath:2019uuo, K:2023gsi, K:2024sla}.
\par 
Lensing anomaly refers to an inconsistency in the estimation of value of lensing parameter from CMB. In particular, this refers to the fact that value of lensing parameter \cite{Calabrese:2008rt} seems to be different from the expected value of one. Lensing anomaly is different from other anomalies, in that effect of lensing is observed at large multipoles. In this work, we would primarily like to investigate the existence of power suppression and explore the connection between power suppression and lensing anomaly from a phenomenological perspective. We shall assume simple extensions to nearly scale invariant primordial power spectrum such as the one which includes running and/or running of running of spectral index and investigate these anomalies. These models are agnostic about power suppression {\it i.e.} the parameters can take values which can lead to power suppression or an enhancement at low multipoles. We perform Bayesian parameter estimation and investigate whether data prefers values which lead to suppression of power at low multipoles. We then investigate the alleviation of lensing anomaly in these models. We discuss the significance of these findings using Akaike and Bayesian information criteria. We forecast the ability of near-perfect fourth generation CMB missions such as ECHO\footnote{\href{https://cmb-bharat.in/}{https://cmb-bharat.in/}} (Exploring Cosmic History and  Origins) to make tighter constraints on model parameters. 

\par 
This paper is organised as follows. We begin with a brief review of essential aspects of standard model of cosmology, CMB data analysis and the anomalies of interest in section \ref{sec:2}. We describe the different analytical templates of primordial power spectrum, that we consider, in section \ref{sec:3}. In section \ref{sec:4}, we perform a Bayesian parameter estimation of these models and investigate power suppression in them. We then explore the connection between power suppression and lensing anomaly in section \ref{sec:5}. In section \ref{sec:6}, we discuss the extent to which these models are preferred by the data using information criteria. In sections \ref{sec:4} to \ref{sec:6}, we work with legacy release of Planck data referred to as Planck Release 3 (PR3) along with some data from baryon acoustic oscillations and some two-point statistics of galaxies.
In \ref{sec:7}, we analyse our models in the light of latest release of Planck data known as Planck Release 4 (PR4). 
We then perform forecast for future CMB mission in section \ref{sec:8}. 
We conclude the paper with a summary and a discussion of our results in section \ref{sec:9}.
\section{Standard model and CMB anomalies}\label{sec:2}
Standard model of cosmology posits a spatially flat Friedmann-Lemaitre-Robertson-Walker geometry for our Universe with inhomogeneities seeded by  Gaussian primordial perturbations \cite{Planck:2018jri, Planck:2019kim}. According to this model, our Universe is mostly composed of cosmological constant ($\Lambda$) and cold dark matter (CDM). Hence, this model is also called as $\Lambda$CDM model \cite{Planck:2018vyg}. 
With just six parameters $\Lambda$CDM model has been able to describe many aspects of our Universe quite well. The six parameters include energy density of baryons ($\Omega_b\,h^2$) and cold dark matter ($\Omega_c\,h^2$) -- which describe the constitution of our Universe, hundred times the angle subtended by the sound horizon on the surface of last scattering ($100\,\theta_{\rm MC}$)-- which describe the amount of expansion since last scattering, reionization depth ($\tau$) -- which informs us about the epoch of reionization and the amount of secondary anisotropies generated in the CMB and two parameters that describe the amplitude ($A_s$) and tilt ($n_s$) of the scalar primordial power spectrum, through the relation 
\begin{equation}
 {\cal P}_s(k)\,=\, A_s \biggl(\frac{k}{k_\star}\biggr)^{n_s\,-\,1}.
\end{equation}
We shall refer to this template for a nearly scale invariant power spectrum as the {\it Standard Ansatz} ($SA$). Such a power spectrum can be very easily generated in a large class of slow roll inflationary models \cite{Planck:2018jri}. 
\par 
Given a theoretical model, we can compute observables, gather data corresponding to these observables and compare both to arrive at information about model parameters. Because of the limitations inherent to the measurement, we cannot arrive at the exact values of the underlying model parameters. Bayes theorem provides a way to estimate the model parameters and to translate the errors in the measurement to our confidence in the estimated value (see, for instance,  \cite{sivia_book}). It can be stated as 
\begin{equation}
 P(\theta|D)\, =\, \frac{P(D|\theta)\,P(\theta)}{P(D)},
\end{equation}
where $P(\theta)$ is our prior knowledge about model parameters $\theta$ before carrying out the observation,  $P(D|\theta)$ is the likelihood that data $D$ will be obtained using the underlying model for the given values of parameters, $P(D)$ is a normalization constant or the evidence and $P(\theta|D)$ is the posterior probability distribution of model parameters. We can use stochastic methods such as Markov Chain Monte Carlo (MCMC) to sample the parameter space and arrive at the posterior probability distribution which captures our degree of belief in the value of model parameters. 
\begin{figure}
 \begin{tabular}{cc}
  \includegraphics[width=0.48\textwidth]{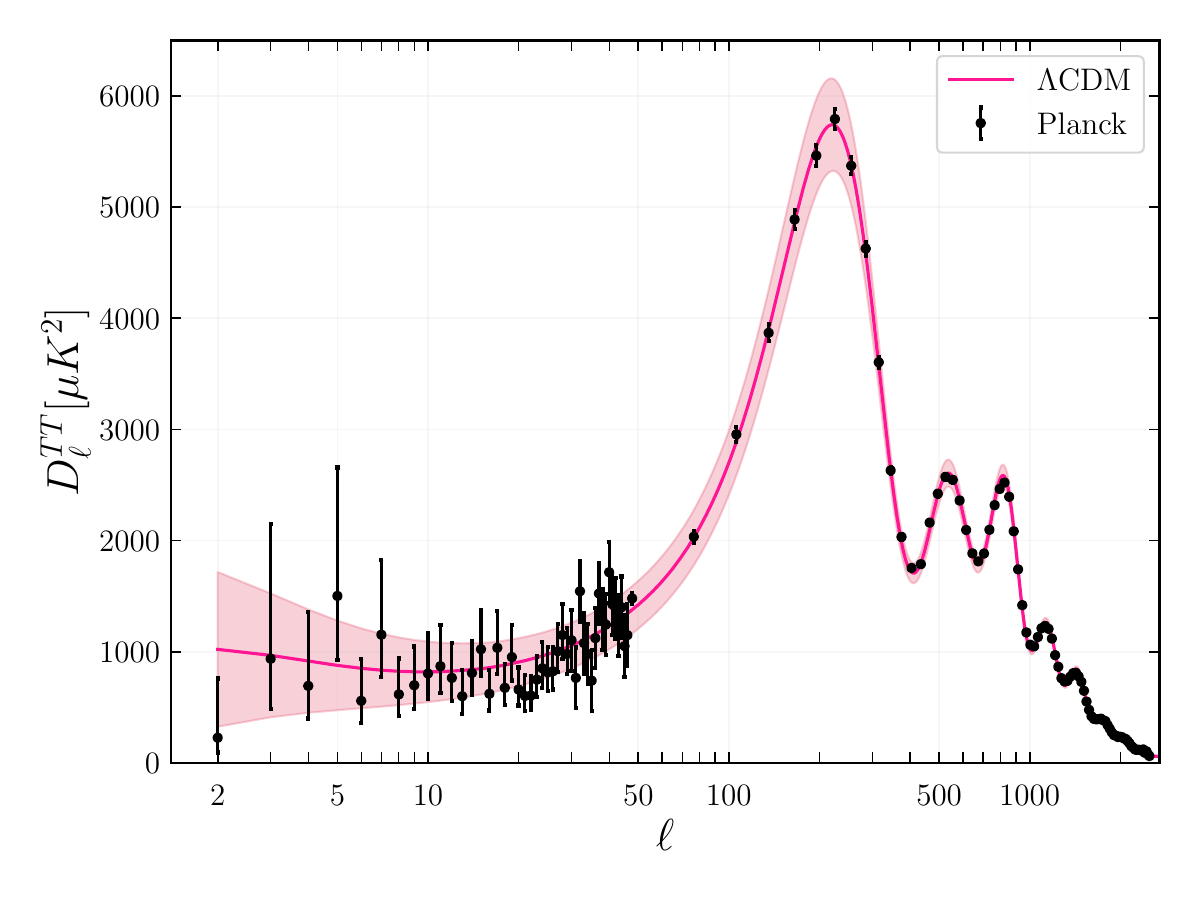}&
  \includegraphics[width=0.48\textwidth]{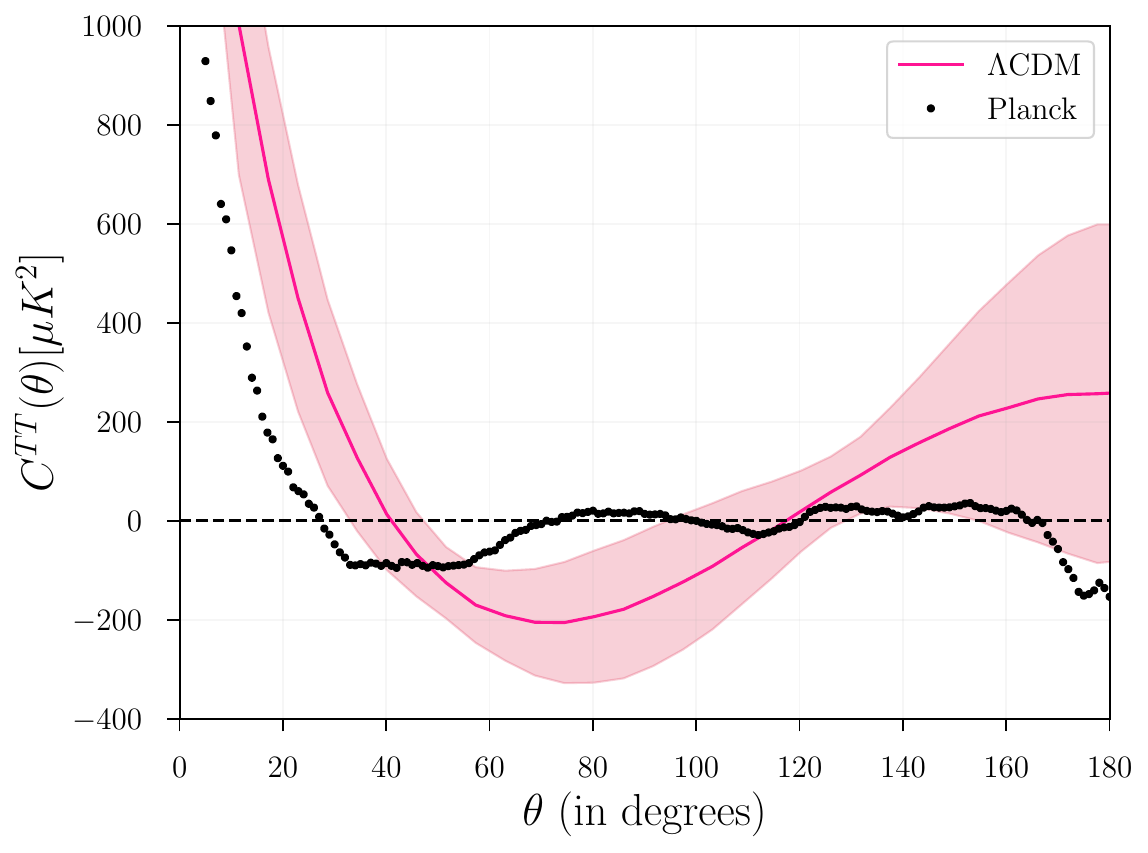}
 \end{tabular}
\caption{\label{fig:Clstd}The plots of $D_\ell^{TT}$ (left) and $C^{TT}(\theta)$ (right) corresponding to the standard model (solid lines) together with data obtained from Planck (black dots)  \cite{Planck:2018nkj}. 
To compute $D_\ell^{TT}$ generated in $\Lambda$CDM model, we worked with marginalised mean values of parameters obtained using Bayesian parameter estimation with PR3 data of temperature, polarisation and lensing obtained by Planck. Power spectrum in angular space $C^{TT}(\theta)$ is computed from $C_\ell^{TT}$ using equation (\ref{eqn:Ctheta}). The data for $C^{TT}(\theta)$ is derived from appropriately masked Planck temperature map  (\texttt{COM\_CMB\_IQU-smica\_2048\_R3.00\_full.fits} together with the common mask \texttt{
COM\_Mask\_CMB-common-Mask-Int\_2048\_R3.00.fits}) using \texttt{healpy} \cite{Gorski:2004by, Zonca:2019vzt}.
In particular, we downgrade the masked map to \texttt{NSIDE} $\,=\, 64$, construct $C_\ell^{TT}$ for $\ell \in [2,\,128]$ and then compute $C^{TT}(\theta)$ using Eqn. (\ref{eqn:Ctheta}).
}
\end{figure}
\par
Because of the mostly linear physics, anisotropies in the CMB provides one of the cleanest data to validate standard model. Hence, most of the current constraints on the standard model has been arrived at using the data from CMB. In this paper we will mainly use state-of-the-art data measured by Planck for our analysis. Anisotropies present in the CMB are temperature fluctuations, electric or E-mode polarisation and magnetic or B-mode polarisation. Since temperature fluctuations and polarisation are random quantities, we can only compare their moments with the corresponding predictions from the model. 
Data of their variance or their power spectra as a function of multipoles can be used to constrain models. We use power spectra of temperature fluctuations ($C_\ell^{TT}$ or TT) and E-mode polarisation ($C_\ell^{EE}$ or EE) and their cross-correlation ($C_\ell^{TE}$ or TE). The B-mode polarisation from a primordial origin, though expected to be created by inflation, has not yet been detected. Lack of B-mode signal in CMB has helped us to rule out certain class of inflationary models. Another observable that has been derived from CMB is the power spectrum of lensing potential ($C_\ell^{\phi\phi}$). Using Boltzmann codes which allow us to compute these observables and using MCMC codes, we can compare a given model with data using Bayesian methods. In this article, we will work with \texttt{CosmoMC} \cite{Lewis:2002ah} and \texttt{Cobaya}\cite{Torrado:2020dgo} together with \texttt{CAMB} \cite{Lewis:1999bs}. We use \texttt{GetDist} \cite{Lewis:2019xzd} to analyse MCMC chains. We use \texttt{CAMB} to compute power spectra.
\par
Figure \ref{fig:Clstd} contains the plot (solid curve) of $D^{TT}_{\ell}\, =\, \ell (\ell + 1)\,C_\ell^{TT}/2\pi$ generated by standard model with parameters set to their marginalised mean values obtained from a Bayesian comparison with Planck legacy data (PR3).
Dots with error bars represent data observed by Planck. These error bars arise due to limitations in the measurement. Cosmic variance is indicated by the shaded region. Unlike other measurement errors which can be minimised by improving the instrument, cosmic variance is an inherent limitation of cosmological measurements (see, for instance, \cite{Weinberg:2008zzc}). 
Cosmic variance of $C_\ell^{XY}$, where $X$ and $Y$ are either T or E,  is $\sqrt{2/(2\,\ell\, +\, 1)}C_\ell^{XY}$. Since cosmic variance is inversely proportional to the number of $m$ values that can be averaged over, it is larger at smaller multipoles. Due to galactic foreground and other point sources, the entire sky may not be observable. To account for this, the above expression is scaled by inverse of the fraction of sky that has been observed, $f_{sky}$, as  $\sqrt{2/[\,(2\,\ell\, +\, 1)f_{sky}\,]}\, C_\ell^{XY}$. 
In figure \ref{fig:Clstd}, we have worked with $f_{sky}\,=\,0.86$ observed by Planck\footnote{Value of $f_{sky}$ varies with choice of foreground mask \cite{Planck:2019nip}.}. As is clear from the figure, the standard model provides a fairly good fit to the Planck data. However, there are some features in the CMB that deviate from predictions of standard model. In the remaining part of this section, we will focus on some of these features. 
\par 
We will mainly consider two anomalies. Firstly, the observed power of temperature fluctuations at low multipoles is lower than that predicted by the standard model. This lack of power is more evident if we plot CMB spectrum in terms of angular separation. The angular power spectrum of temperature fluctuations generated in the standard model is plotted in the right panel of figure \ref{fig:Clstd}. In terms of $C_\ell^{TT}$ it is given by 
\begin{equation}\label{eqn:Ctheta}
C^{TT}(\theta)\, =\, \frac{1}{4\pi}\, \sum_\ell\, (2\,\ell\,+\,1)\, C_\ell^{TT}\, P_\ell(\cos \theta) .
\end{equation}
The shaded region represents the cosmic variance and the black dots represents the observations. From the plot, we can see that while the data for angles greater than $60^0$ is quite close to zero, the solid curve that corresponds to the standard model is not. This lack of power at large angular scales can be quantified using \cite{WMAP:2003elm}
\begin{equation}\label{eqn:Shalf}
 S_{1/2}\,=\, \int_{-1}^{1/2}C(\theta)^2\,{\rm d}(\cos\theta).
\end{equation}
The value of $S_{1/2}$ observed by Planck is $1209$ \cite{Planck:2019evm} compared to $34810$ obtained with the marginalised mean values for standard model.
\par
\begin{figure}[h]
 \begin{tabular}{cc}
  \includegraphics[width=0.48\textwidth]{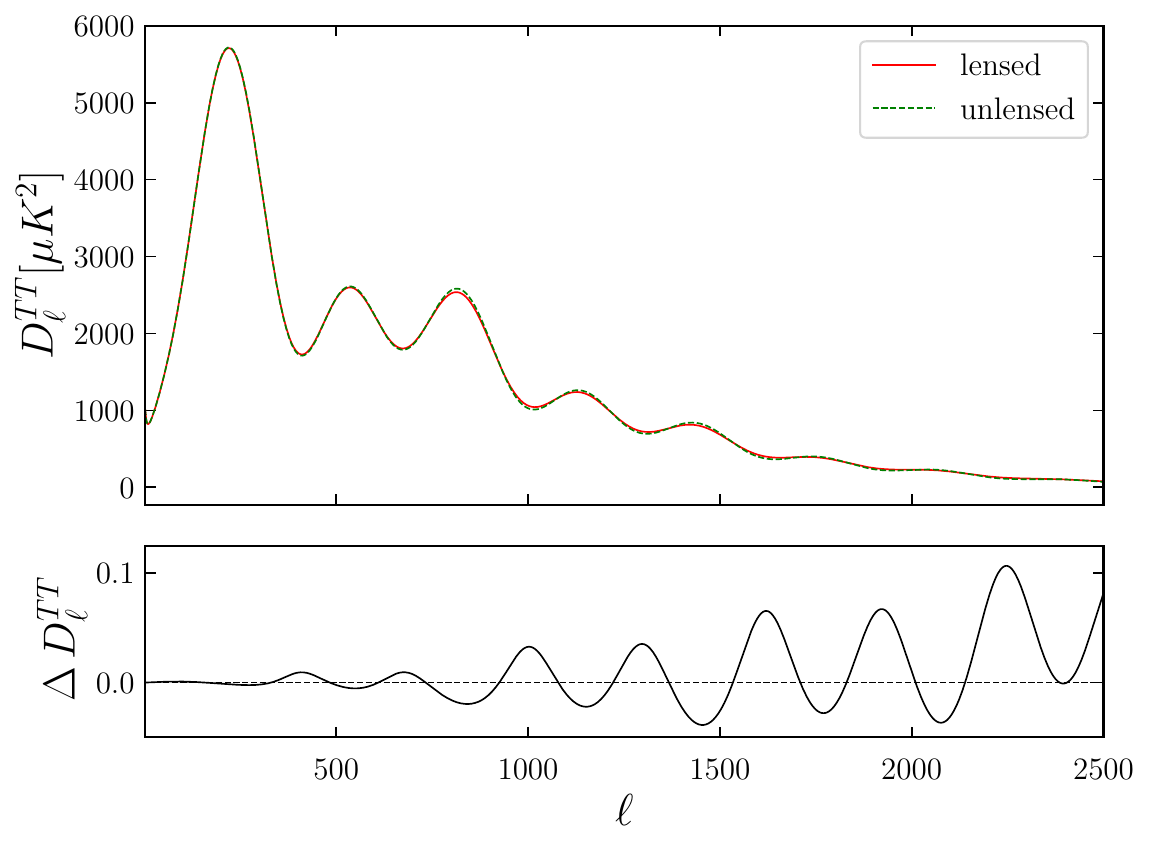}&
  \includegraphics[width=0.48\textwidth]{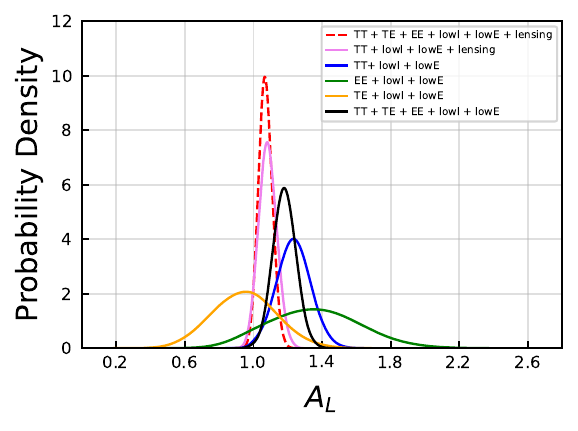}
 \end{tabular}
\caption{\label{fig:lensing}Plots on the left illustrates the effect of lensing on $D_\ell^{TT}$. From the plots, especially from the lower subplot, we see that the effect of lensing is more prominent at higher multipoles.  In the lower subplot, $\Delta\,D^{TT}_{\ell}$ refers to ratio of difference between lensed and unlensed $D^{TT}_{\ell}$ divided by lensed $D^{TT}_{\ell}$. The right panel was obtained by Bayesian estimation of $A_L$ along with the usual six parameters with different combinations of Planck data. We find that value of $A_L$ is closer to one only when \texttt{lensing} data is included.  
}

\end{figure}

CMB as it travels to us get weakly lensed by intervening structure. The effect of this lensing is, see figure \ref{fig:lensing}, a smoothening of acoustic peaks that is primarily felt at large multipoles or low angular scales. 
The effect of the lensing in the spectra of temperature anisotropies and polarisation should be consistent with the spectra of lensing potential. 
To check this consistency, a lensing parameter $A_L$ that multiplies the power spectra of lensing potential was introduced \cite{Calabrese:2008rt}. If we treat $A_L$ as a free parameter and do a Bayesian analysis by comparing the resultant seven parameter model, {\it i.e.} $A_L$ and the usual six parameters, with Planck temperature and polarisation data, the constraints on $A_L$ should be consistent with one. Such a result would imply that we have accounted for the effects of lensing correctly. The results of such an analysis, provided in the right panel of figure \ref{fig:lensing}, indicate otherwise. 
We find that the value of $A_L$ is consistent with one within $2-\sigma$, only if we include data of lensing potential \cite{Planck:2018vyg}. For instance, if we work with Planck legacy TT data alone, then the marginalised mean value of $A_L$ turns out to be $A_L\, =\, 1.24 \pm 0.09$ which is much more than $2-\sigma$ away from the value of one. 
\par
We now proceed to investigate power suppression at low multipoles and its connection with lensing anomaly using certain phenomenological models. We begin by presenting these models in the next section. 

\section{Extensions to standard model}\label{sec:3}
To investigate the connection between primordial physics and CMB anomalies, in addition to the nearly scale invariant form of power spectrum, we will also consider certain extended templates of primordial power spectrum. The remaining four parameters in the $\Lambda$CDM model will be same in all models.
The models that we consider in this work are described below. 

\subsection{Nearly scale invariant model with running of spectral index}
In this model we consider running of the spectral index, {\it i.e.} 
\begin{equation}
    {\cal P_R}(k) = A_s\biggl(\frac{k}{k_{\star}}\biggr)^{n_s-1 + \frac{1}{2}\,\alpha\, ln(k/k_{\star})}.\label{eqn:modelA}
\end{equation}
Including running of spectral index ($\alpha$), this model thus contains seven parameters. 
We will refer to this model as $SA\,+\,\alpha$.
\subsection{Nearly scale invariant model with  running and running of running of spectral index}
In this model, in addition to running of the spectral index, we consider running of running {\it i.e.} 
\begin{equation}
    {\cal P_R}(k) = A_s\biggl(\frac{k}{k_{\star}}\biggr)^{n_s-1 + \frac{1}{2}\,\alpha\, ln(k/k_{\star}) + \frac{1}{6}\, \beta\, ln(k/k_\star)^2}.\label{eqn:modelB}
\end{equation}
Including running of spectral index ($\alpha$) and running of running of spectral index ($\beta$), in addition to the standard model, this model thus contains eight parameters. 
We will refer to this model as $SA\,+\,\alpha\,+\,\beta$.
\subsection{Nearly scale invariant model with running of running of spectral index}
In this model, we consider only running of running of spectral index,   
{\it i.e.} 
\begin{equation}
    {\cal P_R}(k) = A_s\biggl(\frac{k}{k_{\star}}\biggr)^{n_s-1 + \frac{1}{6}\, \beta\, ln(k/k_\star)^2}.\label{eqn:modelC}
\end{equation}
Including running of running of spectral index ($\beta$), in addition to the standard model, this model thus contains seven parameters. 
We will refer to this model as $SA\,+\,\beta$.

\par 
\begin{figure}[h]
 \begin{tabular}{c}
  \includegraphics[width=0.7\textwidth]{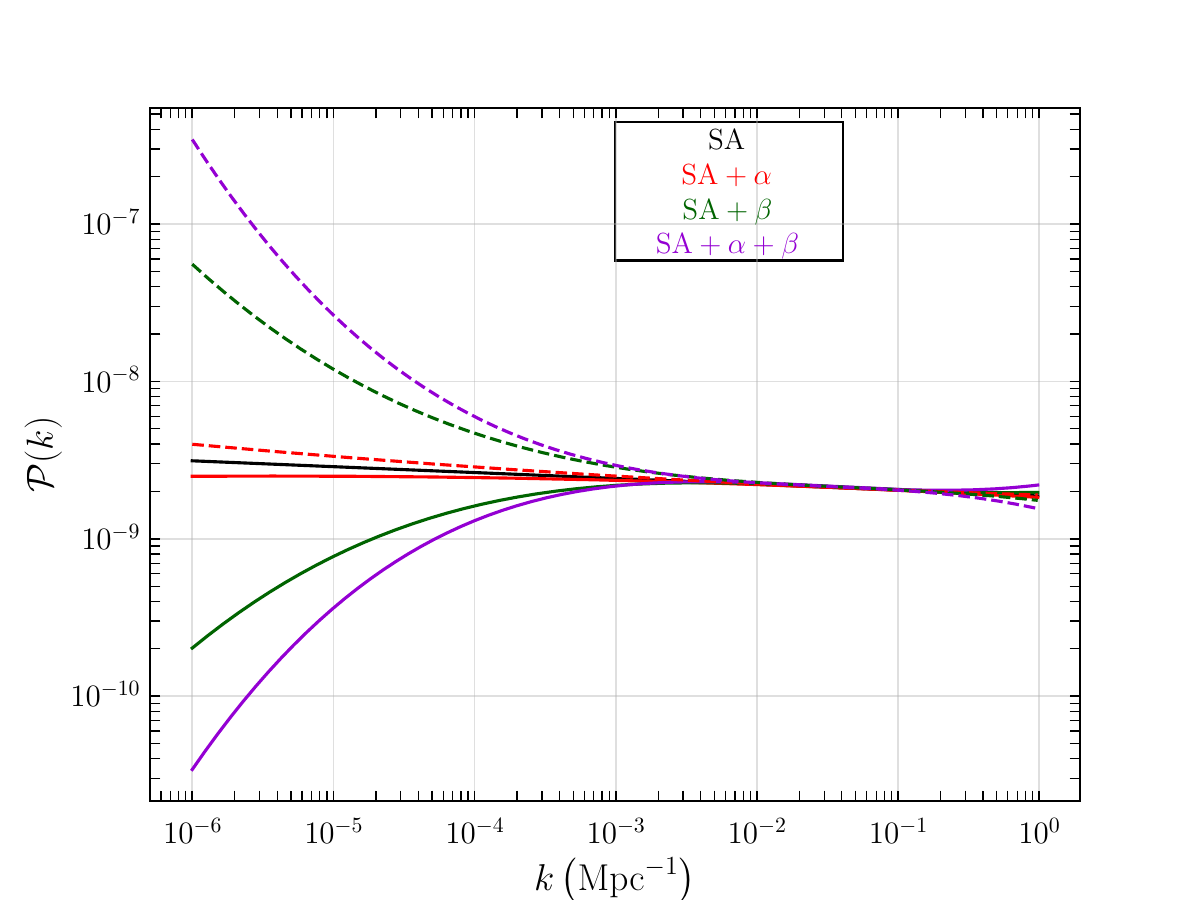}
 \end{tabular}
\caption{\label{fig:primordial} Illustration of power spectra generated in the models that we consider for two different sets of values of model parameters. One set of parameter values lead to power suppression (solid lines) and the other leads to enhancement in power at small wavenumbers (dashed lines). 
To plot solid lines, we have worked with the marginalised mean values of $\alpha$ and $\beta$ given in table \ref{T:A1}. The dashed lines are obtained by working with the same values of $\alpha$ and $\beta$ but with an opposite sign. This plot illustrates that the templates that we consider are agnostic to a suppression in power, {\i.e.}, it can lead to both suppression or enhancement of power at small wavenumbers. }

\end{figure}
Figure \ref{fig:primordial} illustrates the $k$ dependence of different models considered in this paper. As is clear from the figure, depending on the value of the parameters, these models can lead to either a suppression or an enhancement of power at small wavenumbers. Hence, these templates can be used to test whether data indicate the presence of suppression of power. It should be highlighted that, in this analysis, we do not assume that these phenomenological models describe the full behaviour of primordial power spectrum valid over the entire wavelength range. For instance, as is clear from the figure \ref{fig:primordial}, the values of $\alpha$ and $\beta$ in the $SA + \alpha + \beta$ model that favour power suppression at long wavelengths lead to enhancement in power at short wavelengths. However, we are only interested in the ability of this model to test the existence of power suppression at long wavelengths. With this in mind, we will restrict ourselves to CMB data of multipoles $\ell \lesssim 2500$. 
Over this range, these models deviate from $SA$ primarily at low multipoles. 
\section{Power suppression in extensions to standard model}\label{sec:4}
We now employ Bayesian parameter estimation to find out the value of parameters of $SA$, $SA\,+\,\alpha$, $SA\,+\,\alpha\,+\,\beta$ and $SA\,+\,\beta$ models. We will show that the value of model parameters preferred by the data leads to a suppression of power at large angular scales. 
In this section, we have used \verb|CosmoMC| together with \verb|CAMB| for this analysis. In this work, we have assumed a flat prior for all parameters as described in table \ref{T:prior}. 
We have compared these models with a variety of data sets. The most accurate data that we have is that of temperature anisotropies measured by Planck \cite{Planck:2018vyg}. It is in fact one of the best measured observable of our universe. The error bars for this data are better than that due to cosmic variance. 
In this and subsequent two sections, we will work with  PR3 data. Later, in section \ref{sec:7}, we shall extend our investigation to include PR4 data. 
Planck's measurement of temperature fluctuations is available over the  multipoles $\ell = [2,2508]$. 
We use public Planck likelihood \texttt{commander} to analyse low multipole, $\ell \in [2,29]$, TT data (\texttt{lowl}). High multipole temperature data (TT), $\ell \in [30,2508]$, is analysed with \texttt{Plik} likelihood. 
While working with TT, we have also included low $\ell$ EE data (\texttt{lowE}) analysed with \texttt{simall} likelihood to obtain better constraints on reionization depth \cite{Planck:2019nip}.  
The results of Bayesian parameter estimation of our models with \texttt{TT + lowl + lowE} are provided in table \ref{T:A1}.
\begin{table}
    \centering
    \begin{tabular}{|c|c|}
    \hline 
    Parameter & Prior \\
    \hline 
       $\Omega_b\,h^2$  &  [0.005, 0.1]\\
       $\Omega_c\,h^2$  & [0.001, 0.99]\\
       $100\,\theta_{\rm MC}$ & [0.5, 10]\\
       $\tau$ & [0.01, 0.8]\\
       ${\rm{ln}}(10^{10} A_s)$ & [1.61, 3.91]\\
       $n_s$ & [0.8, 1.2]\\
       $\alpha$ & [-1, 1]\\
       $\beta$ & [-1, 1] \\
       $A_L$ & [0, 2.7]\\
       \hline 
    \end{tabular}
    \caption{Table lists the range of values of different parameters, appearing in various models, over which a uniform prior knowledge is assumed. }
    \label{T:prior}
\end{table}

\par 
We also consider the effect of including polarisation data from higher multipoles. 
In particular, along with temperature and low multipole EE spectrum, we consider TE and EE spectra from multipoles $30 - 1996$. We shall refer to this data set as \texttt{TTTEEE + lowl + lowE}. Marginalised mean values and standard deviation of cosmological parameters for various models obtained from their posterior probability distributions are tabulated in table \ref{T:A2}.  
\par 
Finally, we also extend the analysis of our models to include non-CMB data. We include these data with the expectation that additional data on the distribution of matter will lead to better constraints on non-primordial parameters.
In particular, we include measurements of the Baryon Acoustic Oscillations (BAO) data obtained from 6DF \cite{Beutler:2011hx}, MGS \cite{Ross:2014qpa} and SDSS surveys \cite{SDSS:2000hjo}. We also include first year release of data of galaxy clustering and weak lensing from Dark Energy Survey (DES) \cite{DES:2018gui, DES:2018adi, DES:2015wtr}. These include three sets of two-point statistics, namely,  galaxy clustering, galaxy-galaxy lensing and cosmic shear \cite{DES:2017myr}. For brevity, we shall refer to \texttt{TTTEEE + lowl + lowE + BAO + DES} as \texttt{Planck + BAO + DES}. Results of Bayesian parameter estimation upon working with \texttt{TT + lowl + lowE + BAO + DES} and \texttt{Planck + BAO + DES} data sets are given in tables \ref{T:A3} and \ref{T:A4} respectively.
\begin{table}
 \centering 
 \begin{tabular}{|p{1.6cm}|p{2.7cm}|p{2.7cm}|p{2.7cm}|p{2.7cm}|p{2.7cm}|}
 \hline
  & $SA$  & $SA + \alpha$  & $SA + \beta$  & $SA + \alpha$ + $\beta$  \\
  \hline
 $\Omega_b\,h^2$ & $0.0221\,\pm\, 0.00022$ & $0.0222\, \pm\, 0.00023$ & $0.0221\, \pm\, 0.00022$ & $0.022\, \pm\, 0.00027$ \\
  $\Omega_c\,h^2$ & $0.1206\, \pm\,  0.00210$ & $0.1207\,\pm\, 0.00212$ & $0.1216\,\pm\, 0.0022$ & $0.1221\, \pm\, 0.0022$ \\
 $100\,\theta_{\rm MC}$ & $1.0408 \,\pm\, 0.00047$& $1.0408\, \pm\, 0.00047$ & $1.0407\,\pm\, 0.00047$ & $1.0406\, \pm\, 0.00048$\\
  $\tau$ & $0.0517\, \pm\, 0.0080$& $0.0531\, \pm\, 0.0086$ & $0.0554\, \pm\, 0.0087$ & $0.0545\, \pm\, 0.0088$ \\
  ${\rm{ln}}(10^{10} A_s)$ & $3.04 \,\pm\, 0.016$ & $3.04\, \pm\, 0.018$ & $3.05\,\pm\, 0.018$ & $3.05 \,\pm\, 0.019$ \\
  $n_s$ & $0.9627\,\pm\, 0.0057$ & $0.9620 \,\pm\, 0.0059$ & $0.9573\, \pm\, 0.0067$ & $0.9551\,\pm\, 0.0069$\\
  $\alpha$ & 0 & $-0.0040\,\pm\, 0.0076$ & 0 & $0.0136 \,\pm\, 0.0125$ \\
  $\beta$ & 0 & 0 & $0.0133\,\pm\,0.0089$ & $0.0256\,\pm\, 0.0142$ \\
 \hline
 $\chi^2$& $1192.1\,\pm\, 5.5$& $1192.9\,\pm\, 5.7$ & $1191\, \pm\, 5.6$ & $1190.5\,\pm\, 5.8$ \\
 \hline
 \end{tabular}
 \caption{\label{T:A1} Marginalised mean and standard deviation of different model parameters obtained by Bayesian parameter estimation. In this analysis, we have worked with \texttt{TT + low $\ell$ + low E} data. Last row of this table contain the marginalised mean and standard deviation of values of $\chi^2$ obtained from the posterior distribution.}
 \end{table}
%
\begin{table}
 \centering 
 \begin{tabular}{|p{1.6cm}|p{2.7cm}|p{2.7cm}|p{2.7cm}|p{2.7cm}|}
 \hline
  & $SA$  & $SA + \alpha$  & $SA + \beta$  & $SA + \alpha + \beta$  \\
  \hline
 $\Omega_b\,h^2$ & $0.0224\, \pm\, 0.00015$ & $0.0224\, \pm\, 0.00016$ & $0.0224\, \pm\, 0.00015$ & $0.0223\, \pm\, 0.00016$ \\
  $\Omega_c\,h^2$ & $0.1202\,\pm\, 0.00136$ & $0.1204\,\pm\, 0.00140$ & $0.1206\, \pm\,  0.00142$ & $0.1207 \, \pm\, 0.00142$ \\
 $100\,\theta_{\rm MC}$ & $1.0409\, \pm\, 0.00031$ & $1.0409\,\pm\, 0.00032$ & $1.0409\,\pm\, 0.00031$ & $1.0409\, \pm\, 0.00032$ \\
  $\tau$ & $0.0544\, \pm\, 0.0078$ & $0.0559\, \pm\, 0.0081$ & $0.0573\, \pm\,  0.0085$ & $0.0575 \,\pm\, 0.0087$ \\
  ${\rm{ln}}(10^{10} A_s)$ & $3.05\,\pm\, 0.016$ & $3.05 \,\pm\, 0.017$ & $3.05\, \pm\, 0.018$ & $3.05\, \pm\, 0.018$ \\
  $n_s$ & $0.9649\, \pm\, 0.0043$ & $0.9634\, \pm\, 0.0047$ & $0.9614\, \pm\,  0.0051$ & $0.9612\, \pm\, 0.0052$ \\
  $\alpha$ & 0  & $-0.0062\, \pm\, 0.0068 $ & 0  & $0.0011\, \pm\, 0.0102$ \\
  $\beta$ & 0  & 0  & $0.0102\, \pm\, 0.0084$ & $0.0117\, \pm\, 0.0127$ \\
 \hline
 $\chi^2$& $2780.2\, \pm\, 5.8$ & $2780.8\, \pm\, 6.0$ & $2779.8\, \pm\,  5.9$ & $2780.9\, \pm\, 6.1$ \\
 \hline
 \end{tabular}
 \caption{\label{T:A2} Marginalised mean value and standard deviation of cosmological parameters obtained by Bayesian analysis of models with \texttt{TTTEEE + lowl + lowE} data. Mean and standard deviation of $\chi^2$ obtained from the posterior distribution is listed in the last row. }
 \end{table}

\begin{table}
 \centering 
 \begin{tabular}{|p{1.6cm}|p{2.7cm}|p{2.7cm}|p{2.7cm}|p{2.7cm}|p{2.7cm}|}
 \hline
  & $SA$ & $SA + \alpha$ & $SA + \beta$ & $SA + \alpha + \beta$ \\
  \hline
 $\Omega_b\,h^2$ & $0.0223\, \pm\, 0.00019$ & $0.0223 \,\pm\, 0.00021 $ & $0.0223\, \pm\,0.00019  $  & $ 0.0222\, \pm\, 0.00023$ \\
  $\Omega_c\,h^2$ &  $0.1174\, \pm\,0.00101 $& $0.1174 \, \pm\, 0.00101 $ & $0.1175\, \pm\,0.00101$ & $0.1175 \, \pm\, 0.00102$ \\
 $100\,\theta_{\rm MC}$ & $ 1.0411\, \pm\, 0.00041$ & $ 1.0411\,\pm\, 0.00041$ & $ 1.0411\,\pm\,0.00042 $ & $ 1.0411\,\pm\,0.00041 $ \\
  $\tau$ & $ 0.0527\, \pm\, 0.0079 $ & $0.0530\, \pm\, 0.0082$ & $0.0556\,\pm\,0.0086$ & $ 0.0550\, \pm\,0.0087 $ \\
  ${\rm{ln}}(10^{10} A_s)$ & $3.03 \,\pm\, 0.0161$ & $3.03\,\pm\, 0.0173$& $3.04\,\pm\, 0.0175$ & $3.04 \,\pm\, 0.0181$ \\
  $n_s$ & $0.9694 \, \pm\, 0.0040$ & $0.9693\,\pm\, 0.0042$ & $ 0.9669\,\pm \, 0.0046 $ & $0.9656\, \pm\, 0.0047$ \\
  $\alpha$ & 0  & $-0.00064\,\pm\,0.0074$ & 0 & $0.0147 \, \pm\,0.0121 $ \\
  $\beta$ & 0  & 0  & $0.0091\,\pm\, 0.0084$ & $0.0220\, \pm\,0.0137$ \\
 \hline
  $\chi^2$ & $1717.4\, \pm \, 7.4 $ & $ 1718.3\, \pm \,7.5 $ & $1717.2\, \pm 7.5$ & $1716.4\, \pm \, 7.8$\\
  \hline
 \end{tabular}
 \caption{ \label{T:A3}
 Same as Table \ref{T:A2} but with \texttt{TT + lowl + lowE + BAO + DES} data.}
 
 \end{table}
\begin{table}
 \centering 
 \begin{tabular}{|p{1.6cm}|p{2.7cm}|p{2.7cm}|p{2.7cm}|p{2.7cm}|p{2.7cm}|}
 \hline
  & $SA$ & $SA + \alpha$  & $SA + \beta$  & $SA + \alpha + \beta$  \\
  \hline
 $\Omega_b\,h^2$ & $0.0225\, \pm\, 0.00013$ & $0.0225\,\pm\, 0.00014$ & $0.0225\, \pm\,  0.00013$  & $0.0225\, \pm\, 0.00014$ \\
  $\Omega_c\,h^2$ &  $0.1180\, \pm\, 0.00087$& $0.1180\, \pm\,  0.00087$ & $0.1181\, \pm\, 0.00090$ & $0.1181\, \pm\, 0.00088$ \\
 $100\,\theta_{\rm MC}$ & $1.0411\, \pm\, 0.00029$ & $1.0411\,\pm\, 0.00029$ & $1.0411\,\pm\, 0.00029$ & $1.0411\,\pm\, 0.00029$ \\
  $\tau$ & $0.0541\,\pm\, 0.0078$ & $0.0548\, \pm\, 0.0079$ & $0.0562\,\pm\, 0.0083$ & $0.0565\, \pm\, 0.0083$ \\
  ${\rm{ln}}(10^{10} A_s)$ & $3.04 \,\pm\, 0.016$ & $3.04\,\pm\, 0.016$& $3.04\,\pm\, 0.017$ & $3.04\,\pm\, 0.017$ \\
  $n_s$ & $0.9693\, \pm\, 0.0037$ & $0.9688\,\pm\, 0.0039$ & $0.9674\,\pm\, 0.0043$ & $0.9671\, \pm\, 0.0043$ \\
  $\alpha$ & 0  & $-0.0025\,\pm\, 0.0067$ & 0 & $0.0045\, \pm\, 0.0102$ \\
  $\beta$ & 0  & 0  & $0.0069\,\pm\, 0.0083$ & $0.0113\, \pm\, 0.0125$ \\
 \hline
  $\chi^2$ & $3306.1 \pm 7.7$  & $3306.9 \pm 7.9$ & $3306.3 \pm 7.9$&  $3307.0 \pm 8.1$\\
  \hline
 \end{tabular}
 \caption{\label{T:A4} 
 Same as table \ref{T:A2} but with \texttt{Planck + BAO + DES} data.
 }
 \end{table}
\begin{figure}
\begin{tabular}{cc}
\includegraphics[width=0.48\textwidth]{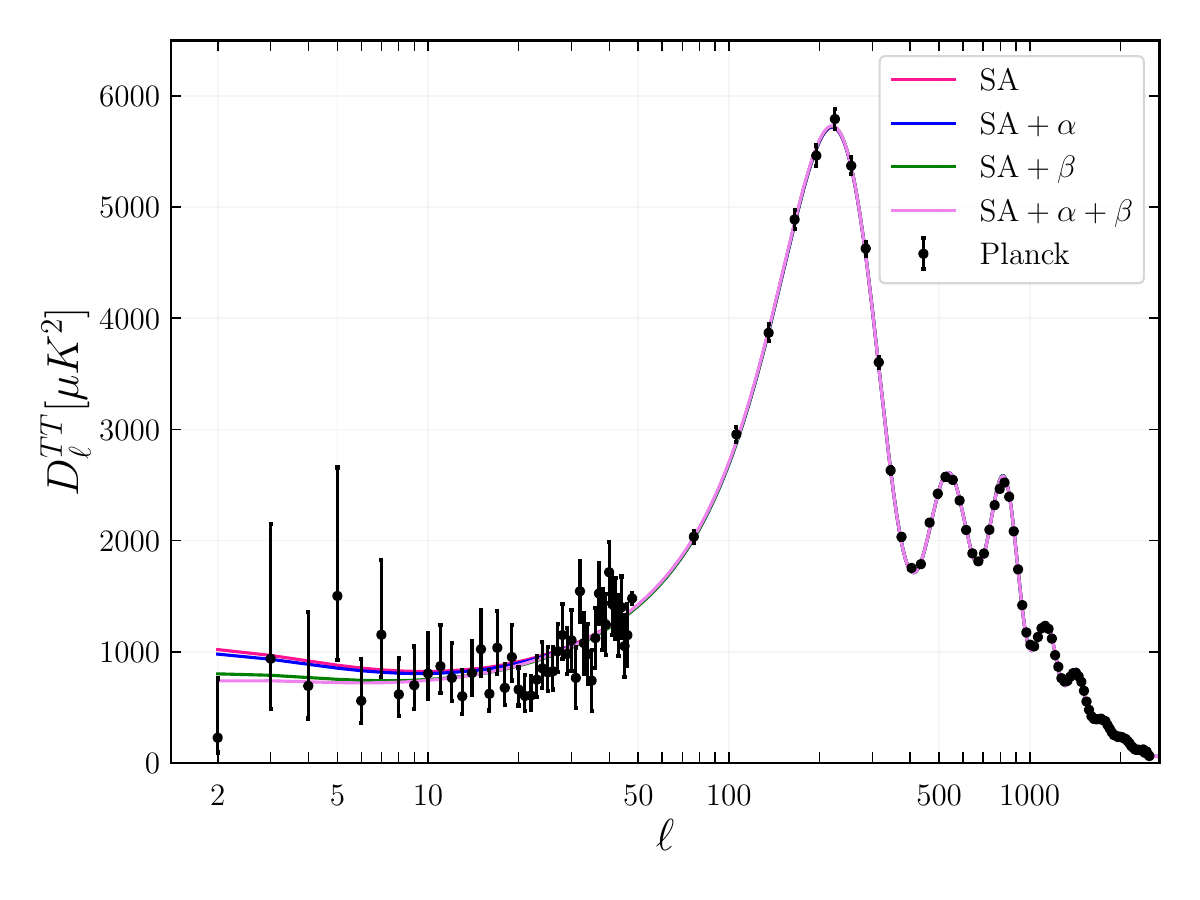}&
\includegraphics[width=0.48\textwidth]{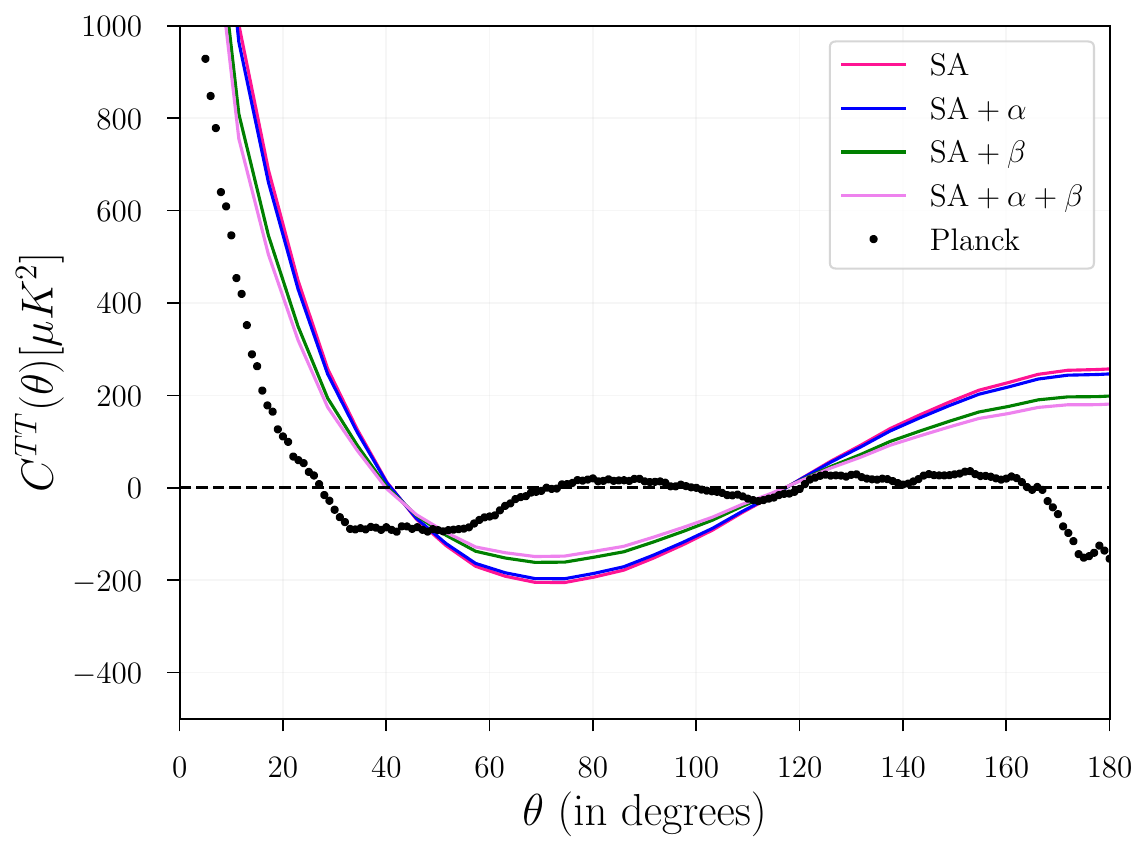}
\end{tabular}
\caption{\label{fig:ClCtheta} Plots of $D_\ell^{TT}$ as a function of multipoles (left) and of $C^{TT}(\theta)$ (right) corresponding to different models are given. 
We have worked with marginalised mean values of parameters given in table \ref{T:A1} which were obtained by comparing models with \texttt{TT + lowl + lowE} data. 
Figure illustrates that data prefers parameter values that lead to suppression in power at low multipoles or at large angles. }
\end{figure}
\par 
On the left panel of figure \ref{fig:ClCtheta}, $D_{\ell}^{TT}$ is plotted as a function of multipoles. In obtaining this plot, we have worked with the marginalised mean values of cosmological parameters obtained by comparing models with \texttt{TT + lowl + lowE} data. Figure illustrates that power at lower multipoles for the extended models is lower in amplitude compared to that generated in $SA$. This is much more evident in the plot of corresponding angular power spectra, Eqn. (\ref{eqn:Ctheta}), plotted on the right side of the same figure. Figure illustrates that at angular scales larger than $60^0$, power is lesser for the extended models. The lack of power at large angular scales is quantified by the $S_{1/2}$, see Eqn. (\ref{eqn:Shalf}), values listed in table \ref{T:Shalf}. Unlike figure \ref{fig:ClCtheta}, table not only lists the values of $S_{1/2}$ corresponding to the marginalised mean values derived from \texttt{TT + lowl + lowE} data given in table \ref{T:A1}, but also shows the values corresponding to analysis performed with other data sets listed in tables \ref{T:A2}, \ref{T:A3} and \ref{T:A4}. Note that lower value of $S_{1/2}$ implies lower power at large angular scales. Table indicates that for all extensions to standard model considered in this paper, data chooses marginalised mean value of parameters 
that lead to a lowering of $S_{1/2}$ or equivalently leads to power suppression at large angular scales. This is the first result of this paper. We note that power suppression is largest for $SA + \alpha + \beta$ model. The $SA + \beta$ model also lead to a comparable, but slightly higher, value of $S_{1/2}$. The value of $S_{1/2}$ for $SA + \alpha$ is only slightly smaller than that generated in $SA$. It is interesting to note that when polarisation data is considered the value of $S_{1/2}$ increases in all models except $SA + \alpha$. On the other hand, when \texttt{BAO} and \texttt{DES} data are considered in addition to \texttt{TT + lowl + lowE} data, value of $S_{1/2}$ increases for all models except for the case of $SA$. A similar behaviour is observed when BAO and DES are added to \texttt{Planck} data.


\begin{table}
    \begin{tabular}{p{5cm} p{2cm} p{2cm} p{2cm} p{2cm}}
    \hline
    & $SA$ & $SA + \alpha$ & $SA + \beta$  & $SA + \alpha + \beta$ \\
  \hline
  \texttt{TT + lowl + lowE} & 34646 & 31878 &  21064 &  17699\\
  &&&&\\
  \texttt{TTTEEE + lowl + lowE}& 34834 & 30640 & 23721 & 22990\\
  &&&&\\
  \texttt{TT + lowl + lowE + BAO + DES}& 34336 & 33907 & 24611 &  20795\\
  &&&&\\
  \texttt{Planck + BAO + DES}& 34513 & 32743 & 26600 & 24909 \\
  &&&&\\
  \hline
    \end{tabular}
    \caption{\label{T:Shalf} Values of $S_{1/2}$ corresponding to different models are tabulated in different columns of this table. $S_{1/2}$ was obtained by evaluating Eqn. (\ref{eqn:Shalf}). Different rows correspond to marginalised mean values of model parameters obtained with different data sets as given in tables \ref{T:A1}, \ref{T:A2}, \ref{T:A3} and \ref{T:A4}. }
\end{table}

\section{Connection between power suppression and lensing anomaly }\label{sec:5}
We now investigate the connection between suppression of power at large angular scales and the lensing anomaly. In earlier works, it was shown, in the context of loop quantum cosmology, that a primordial power spectra with a lack of power at long wavelengths could alleviate lensing anomaly \cite{ Ashtekar:2021izi, Ashtekar:2020gec, Martin-Benito:2023nky}. In this section, we investigate the extent to which lensing anomaly gets resolved in these extensions to standard model. We perform two types of analysis.
\begin{figure}
     
        \centering
        \begin{tabular}{cc}
        \includegraphics[width=0.48\linewidth]{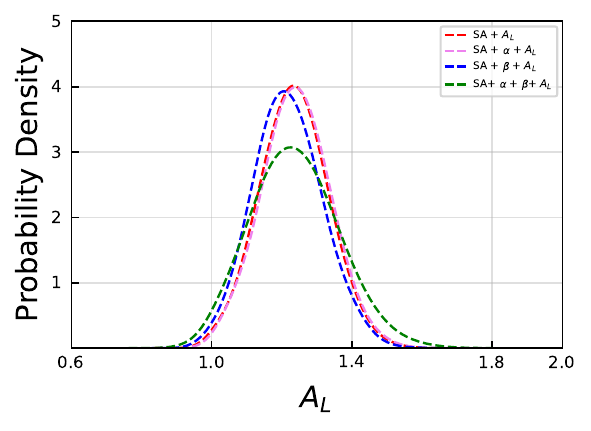}&
        \includegraphics[width=0.48\linewidth]{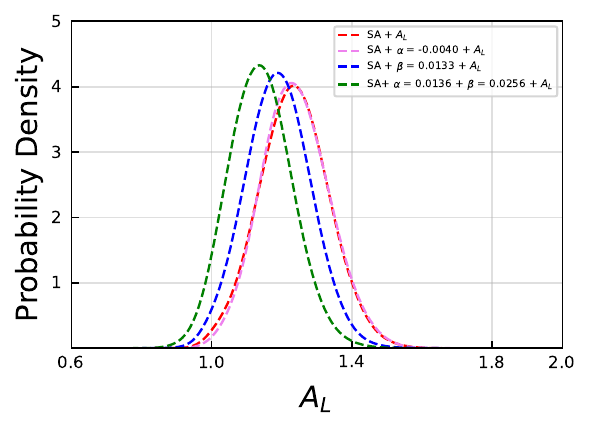}
        \end{tabular}
        \caption{\label{fig:AL} Plot of marginalised probability distribution of $A_L$ for all four models obtained by following Method I (left) and Method II (right). 
         We have worked with \texttt{TT + lowl + lowE} data.
         In Method I, all model parameters including $\alpha$ and $\beta$ are varied, whereas in Method II, we have fixed the values of $\alpha$ and $\beta$ to their respective mean values given in table \ref{T:A1}. We find from Method I that the probability distribution for $A_L$ becomes broader for models $SA\, +\, \beta$ and $SA + \alpha + \beta$ and hence $A_L\,=1$ becomes more probable. From Method II, we find that the probability distribution indeed shifts closer to one for models $SA\, +\, \beta$ and $SA + \alpha + \beta$ which leads to suppression of power at low multipoles. 
        }
\end{figure}
\begin{table}
    \begin{tabular}{p{3cm} p{3cm} p{3cm} p{3cm} p{3cm}}
    \hline
    & $SA$ & $SA + \alpha$ & $SA + \beta$  & $SA + \alpha + \beta$ \\
  \hline
  Method I & $1.240\, \pm\, 0.094$ & $1.246\,\pm 0.096$ & $1.221\,\pm\, 0.100$ & $1.242\,\pm\,0.126$\\
  &&&&\\
  Method II & - & $1.242\,\pm\, 0.096$ & $1.194\,\pm\,0.091$ & $1.142\,\pm\,0.091$\\
  &&&&\\
  \hline
    \end{tabular}
    \caption{\label{T:AL} Marginalised mean and standard deviation of $A_L$ corresponding to different models are tabulated. We have worked with \texttt{TT + lowl + lowE} data. 
    Both methods show that lensing anomaly is alleviated for models $SA + \beta$ and $SA + \alpha + \beta$. 
    From first row of the table, we see that though mean value of $A_L$ of different models remains largely unchanged, their standard deviation becomes larger. In this way, the value of $A_L\,=\,1$ become more probable.  From the second row, we find that mean value of $A_L$ decreases and becomes closer to one, thus alleviating the lensing anomaly. }
\end{table}
\begin{figure}
     
        \centering
        \begin{tabular}{c}
        \includegraphics[width=0.7\linewidth]{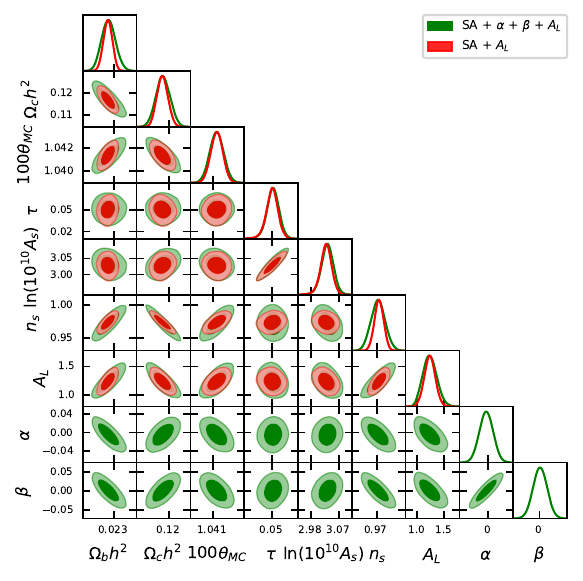}
        \end{tabular}
        \caption{\label{fig:contour_method1} Figure compares marginalised confidence contours between different pairs of cosmological parameters in $SA$ (red) and $SA\,+\,\alpha\,+\,\beta$ (green) model obtained with \texttt{TT + lowl + lowE} data. We have treated $A_L$ as a free parameter in this analysis.  
        We find that the confidence contours of model with $\alpha$ and $\beta$ are broader. This brodening alleviates the lensing anomaly observed in Method I, as seen in the left panel of figure \ref{fig:AL} and the first row of table \ref{T:AL}.}
        
\end{figure}
\par 
In first approach, referred to as Method I, we perform Bayesian parameter estimation for the three extensions to standard model after including the lensing parameter ($A_L$) as a free parameter. This extends the number of free parameters in each model by one.  
We have worked with \texttt{TT + lowl + lowE} data. 
Marginalised posterior probability distribution of the lensing parameter $A_L$, obtained using this approach is plotted in the left side of figure \ref{fig:AL}. We can see from the figure that the probability distribution becomes broader for $SA + \beta$ and $SA + \alpha + \beta$ model. Thus, the value of $A_L = 1$ become more probable. This point is more clear if we look at the table \ref{T:AL} which tabulates the mean and standard deviation of $A_L$ for different models. Note that, in $SA + \alpha + \beta$ model, though mean value $A_L\,=\,1$ remains the same as that in $SA$, its standard deviation is larger, making the value of $A_L\,=\,1$ within $2-\sigma$. In the case of $SA + \beta$ model, the mean value of $A_L$ is slightly lower and the standard deviation is slightly larger making the value of $A_L\,=\,1$ more probable. The constraints on $A_L$ in both $SA$ and $SA + \alpha$ models are largely similar. 
\begin{figure}
     
        \centering
        \begin{tabular}{c}
        \includegraphics[width=0.7\linewidth]{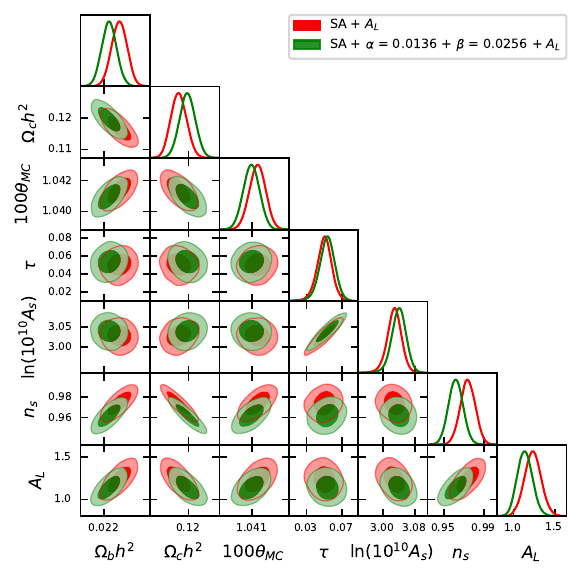}
        \end{tabular}
        \caption{\label{fig:contour_method2} 
        Same as figure \ref{fig:contour_method1}, but for Method II. 
        Unlike Method I, both models have seven parameters, {\it viz.} six standard parameters and the lensing parameter. Compared to $SA$ model, $SA\,+\,\alpha\,+\,\beta$ leads to lesser power at low multipoles. We find that in the latter model, confidence contours of all six cosmological parameters shift in such a way that $A_L\,=\,1$ becomes more probable. This is consistent with the results shown in the right panel of figure \ref{fig:AL} and the second row of table \ref{T:AL}.}
        
\end{figure}
\par 
In the second approach, referred to as Method II, we fix values of $\alpha$ and $\beta$ to those given in table \ref{T:A1} and then vary the remaining parameters of the model along with lensing parameter. The marginalised mean and standard deviation of the lensing parameter is given in the second row of table \ref{T:AL}. The mean value of lensing parameter obtained in the $SA + \alpha + \beta$ model is considerably smaller than that obtained for $SA$. This brings the value $A_L = 1$ well within $2-\sigma$ range. We also find that the value of $A_L$ is lower for $SA + \beta$  model making the value of $A_L\,=\,1$ more probable. As in the first method, we find that constraints on $A_L$ in both $SA$ and $SA + \alpha$ models are largely similar. 
To summarize, both methods show that the extended models which lead to power suppression also makes the value of $A_L = 1$ more probable and thus making it less anomalous. This is the second result of this paper. 
\par 
To understand the reason for the alleviation of lensing anomaly in both approaches, we compare the marginalised confidence contours of all model parameters.  We shall focus on  $SA$ and $SA\, +\, \alpha\, +\, \beta$ models when $A_L$ is considered as a free parameter. Figures \ref{fig:contour_method1} and \ref{fig:contour_method2} provide the plots obtained using Method I and Method II, respectively. From figure \ref{fig:contour_method1}, we see that the joint confidence contours of all pairs of parameters are broader in the case of $SA\, +\, \alpha\, +\, \beta$ model. 
\newvchanges{The parameters $\alpha$ and $\beta$ are significantly correlated with four of the six cosmological parameters of the standard $SA$ model. Further,  parameters $\alpha$ and $\beta$ are negatively correlated with the lensing parameter $A_L$. Hence, positive values of $\alpha$ and $\beta$, which lead to power suppression, will result in a decrease of the lensing parameter $A_L$. This is clearly evident from
Figure \ref{fig:contour_method2}, which plots the joint confidence contours obtained using Method II.} We find that when we fix $\alpha$ and $\beta$ to the value favoured by \texttt{TT + lowl + lowE} data, contours of the standard six cosmological parameters shifts. Unlike figure \ref{fig:contour_method1}, the widths of confidence contours of both models are largely comparable. \newvchanges{Owing to the negative correlation of the parameters $\alpha$ and $\beta$ with the $A_L$, the shift in the contours decreases the mean value of $A_L$ and makes it closer to one.} Thus, Method I shows that an alleviation of lensing anomaly in the $SA\, +\, \alpha\, +\, \beta$ model arises through a broadening of contours. On the other hand, Method II shows that, when there is power suppression (obtained when $\alpha$ and $\beta$ are fixed), confidence contours of all six cosmological
parameters shift in such a way that $A_L$ = 1 becomes more probable.
\newvchanges{It should be noted that Method II is only the distribution of parameters conditioned on the value of $\alpha$ and $\beta$ fixed to their marginalised mean values, whereas Method I represents the full marginalised constraints .}

\par 
It is interesting to ask what happens if we include additional data in our analysis. To answer this, we shall focus on models $SA + \beta$ and $SA + \alpha + \beta$ and work with data sets \texttt{TTTEEE + lowl + lowE}, \texttt{TT + lowl + lowE + BAO + DES} and \texttt{Planck + BAO + DES}. 
The results of this analysis are given in table \ref{T:ALplus}. 
\par 
When we consider additional data sets, we observe that the mean values of $A_L$ as well as its error bars are smaller. When we include CMB polarisation data, we note that the mean values of $A_L$ obtained for $SA + \beta$ and $SA + \alpha + \beta$ are lower than that obtained for $SA$. This is true regardless of whether we fix the values of $\alpha$ and $\beta$ to their marginalised mean values or not. However alleviation of lensing anomaly is more in the case where $\alpha$ and $\beta$ are fixed as per table \ref{T:A2}. 
When we consider the data set \texttt{TT + lowl + lowE + BAO + DES}, we observe that BAO and DES data sets do not affect the mean value of $A_L$. This is evident from a comparison of mean values of $A_L$ given in tables \ref{T:AL} and \ref{T:ALplus} for the case of $SA$. However inclusion of BAO and DES reduces the error bars causing a worsening of lensing anomaly. The expected value of one for the lensing parameter is more than $3-\sigma$ away for $SA$. Analysis using Method I shows that  $A_L = 1$ comes within $3-\sigma$ for $SA + \alpha + \beta$ model. This is also observed in analysis using Method II. Though the mean value of $A_L$ observed for $SA+\beta$ model is slightly lower than $SA$, the alleviation of lensing anomaly is lesser  than that observed in $SA + \alpha + \beta$. We also consider a combination of data sets used in the previous two analysis, {\it i.e.} \texttt{Planck + BAO + DES}. Recall that \texttt{Planck} refers to high-$\ell$ \texttt{TTTEEE} along with \texttt{lowl + lowE}. Though there is a lowering of mean value of $A_L$, it is minimal. Consistent with rest of the analysis, lowering of $A_L$ is observed most in the $SA + \alpha + \beta$ model when $\alpha$ and $\beta$ are fixed to their marginalised mean values. This analysis shows that the alleviation of lensing anomaly weakens when additional data is included. The alleviation of lensing anomaly is best observed in analysis of \texttt{TT + lowl + lowE} data. It is interesting to compare this with the observations of table \ref{T:Shalf} which shows that $S_{1/2}$ value and hence suppression is most evident in analysis with \texttt{TT + lowl + lowE} data. Thus, power suppression is indeed connected to alleviation of lensing anomaly. Larger the suppression, better is the alleviation of lensing anomaly. At the close of this section it should be remarked that E mode polarisation and data from large scale structure have not been measured as well as the temperature anisotropy. So, a higher value of $S_{1/2}$ and hence a lesser amount of alleviation of lensing anomaly favoured by these additional data do not necessarily indicate absence of these effects. A more refined measurement, particularly of E mode polarisation may be needed to make a more concrete statement about power suppression at large angular scales. A better treatment of foreground and other instrumental systematics has resulted in an upgraded version of CMB likelihood. We will investigate power suppression and lensing anomaly using these new likelihoods in section \ref{sec:7}. Furthermore, several near ultimate future CMB missions have been proposed. We will make forecasts for such a mission in section \ref{sec:8}. Before we proceed to these investigations, it is prudent to study the significance of our findings. 
\begin{table}
    \begin{tabular}{p{3cm} p{3cm} p{3cm} p{3cm} p{3cm}}
    \hline
    &  & $SA$ & $SA + \beta$  & $SA + \alpha + \beta$ \\
    \hline 
    \multirow{4}{70pt}{\texttt{TTTEEE + lowl + lowE}} & Method I & $1.183\, \pm\, 0.067$& $1.174 \, \pm \, 0.068$  & $ 1.180\,\pm\,0.070$\\
    &&&&\\
    & Method II & - &$1.165\,\pm\,0.066$ & $ 1.161\,\pm\, 0.066$\\
    &&&&\\
    \hline
    \multirow{4}{70pt}{\texttt{TT + lowl + lowE + BAO + DES}} & Method I & $1.240 \, \pm \, 0.075$& $1.233\,\pm\,0.076$ & $1.226\,\pm\,0.080$\\
    &&&&\\
    & Method II & - & $1.231 \,\pm\, 0.075$ & $1.203\,\pm\, 0.074$\\
    &&&&\\
    \hline
    \multirow{4}{70pt}{\texttt{Planck + BAO + DES}} & Method I & $1.204\, \pm \,0.060$ &  $1.203\, \pm \,0.060$ & $1.199\,\pm\,0.061$\\
    &&&&\\
    & Method II & - & $1.200\,\pm\,0.060$ & $1.195\,\pm\,0.060$\\
    &&&&\\
    \hline
\end{tabular}
\caption{\label{T:ALplus} Marginalised mean and standard deviation of $A_L$ corresponding to different models when compared with additional data sets are tabulated. As in previous table results obtained using two different methods are given.}
\end{table}
\section{Significance of results}\label{sec:6}
A relevant question in this analysis concerns the significance of lack of power on large scales to warrant the addition of extra parameters such as $\alpha$ and $\beta$.
To study the significance with which the extensions to the standard model are favoured, we consider two criteria namely the Akaike Information Criterion (AIC) \cite{Akaike:1974vps} and Bayesian Information Criterion (BIC) \cite{Schwarz:1978tpv}. AIC and BIC are defined as follows,
\begin{eqnarray}
{\rm AIC}\, &=&\, 2\,n\,-\,2\,\ln\, \mathcal{L},\label{eqn:AIC}\\
{\rm BIC}\, &=&\, n\,\ln\, N\, -\,2\,\ln\, \mathcal{L},\label{eqn:BIC}
\end{eqnarray}
where $n$ is the total number of free parameters in a given model, $N$ is the number of data points and $\mathcal{L}$ is the maximum value of likelihood function for a given choice of model and data. The above definitions indicate that BIC penalise addition of extra parameters much more than AIC. We obtain the maximum likelihood required for computing AIC and BIC using Powell's bounded minimization routine (\texttt{action = 2}) of \verb|CosmoMC|. 
For these analyses, we assume a flat prior as described in table \ref{T:prior}. We work with $SA$ as the reference model. Since, power suppression and hence alleviation of lensing anomaly is more observed in $SA + \beta$ and $SA + \alpha + \beta$ models, we restrict our attention to them. As in the previous sections, we work with various combination of data sets. 
We note that \texttt{TT + lowl + lowE} contains 2535, \texttt{TTTEEE + lowl + lowE} contains 6469, \texttt{TT + lowl + lowE + BAO + DES} contains 3443 and \texttt{Planck + BAO + DES} contains 7377 data points respectively. The number of model parameters includes nuisance parameters. This is different for each data set. They are 21, 27, 41 and 47 for $SA$  when working with \texttt{TT + lowl + lowE}, \texttt{TTTEEE + lowl + lowE}, \texttt{TT + lowl + lowE + BAO + DES} and \texttt{Planck + BAO + DES} respectively. Models $SA + \beta$ and $SA + \alpha + \beta$ has one and two extra parameters, respectively. The results of this analysis are given in table \ref{T:IC}. 
\begin{table}
    \begin{tabular}{p{3cm}p{3cm}p{3cm}p{3cm}p{3cm}}
    \hline
    Data & Model & $\chi^2$ & $\Delta$ AIC & $\Delta$ BIC\\
    &&&&\\
    \hline
    \multirow{6}{70pt}{\texttt{TT + lowl + lowE}} & $SA$ & 1180.1202 & 0 & 0\\
    &&&&\\
    & $SA + \beta$ &  1178.4433 & 0.3230 & 6.161\\
    &&&&\\
    & $SA + \alpha + \beta$ & 1176.7702 & 0.649 & 12.325 \\
    &&&&\\
    \hline
    \multirow{6}{70pt}{\texttt{TTTEEE + lowl + lowE}} & $SA$ & 2766.0492 & 0 & 0\\
    &&&&\\
    & $SA + \beta$ &  2764.8400 & 0.7908 & 7.565\\
    &&&&\\
    & $SA + \alpha + \beta$ & 2764.4662 & 2.416 & 15.966\\
    &&&&\\
    \hline
    \multirow{6}{70pt}{\texttt{TT + lowl + lowE + BAO + DES}} & $SA$ & 1699.2963 & 0 & 0\\
    &&&&\\
    & $SA + \beta$ & 1697.6208 & 0.324 & 6.468\\
    &&&&\\
    & $SA + \alpha + \beta$ & 1695.5091 & 0.212& 12.501\\
    &&&&\\
    \hline
    \multirow{6}{70pt}{\texttt{Planck + BAO + DES}} & $SA$ & 3285.5571 & 0 & 0\\
    &&&&\\
    & $SA + \beta$ &  3284.3935& 0.836 & 7.742\\
    &&&&\\
    & $SA + \alpha + \beta$ & 3284.3009 & 2.743& 16.555\\
    &&&&\\
    \hline
    \end{tabular}
    \caption{\label{T:IC} Table of $\Delta$AIC and $\Delta$BIC of $SA$, $SA + \beta$ and $SA + \alpha + \beta$ computed using formulae \ref{eqn:AIC}, \ref{eqn:BIC}. We have set $SA$ as the reference model. $\chi^2$ was estimated using Powell's bounded minimization scheme. In general, $\Delta$AIC or $\Delta$BIC between [0,2] indicate weak preference for $SA$, value between (2,6] indicate moderate preference for $SA$ and value greater than 6 indicate strong preference. }
\end{table}
\par 
According to these information criteria \cite{Capozziello:2020ctn, Hu:2022udt}
if the difference in AIC (BIC) of a model being tested with respect to the reference model, $\Delta$AIC ($\Delta$BIC), is between $[0,2]$ then it indicates a weak preference for the reference model. If $\Delta$AIC ($\Delta$BIC) is between $(2,6]$ then it indicates a moderate preference for the reference model. A $\Delta$AIC ($\Delta$BIC) greater than $6$, indicates a strong preference for the reference model. 
The table \ref{T:IC} lists $\chi^2$ corresponding to the best fit value for different models obtained upon comparison with various data sets and values of $\Delta$AIC and $\Delta$BIC derived from it. 
Our calculations of $\Delta$AIC indicate a weak preference for $SA$ compared to its extensions when polarisation data is not included. When polarisation is included, $SA$ becomes moderately favoured compared to $SA + \alpha + \beta$ model. Compared to the $SA + \beta$ model, the preference for $SA$ remains weak even when high-$\ell$ polarisation data is included. 
The addition of BAO or DES does not seem to considerably affect our estimates of AIC. This indicates that BAO or DES data do not constrain $\alpha$ or $\beta$ much. 
\par 
According to BIC criterion, data strongly favours $SA$ model when working with \texttt{TT + lowl + lowE} data. 
The addition of high-$\ell$ polarisation data improves the preference. However, as in the case of AIC, addition of BAO or DES data sets does not improve the preference much. The strong preference for $SA$ when using BIC can be attributed to its dependence on the number of data points. BIC penalises addition of parameters through the term $n \ln N$ compared to $2\,n$ for AIC. This implies that larger the number of data points higher will be the penalty for adding extra parameters. This would be understandable if we were introducing extra parameters to describe a feature which is present in all the data points. However, the feature of our interest is a lack of power that is limited to low multipoles. Hence, while considering BIC to select models which differ by a local feature, it is more reasonable to work with data points which are centered around the feature \footnote{To find out whether a cobra has eaten a frog, we just need to look at its belly.}. In table \ref{T:lowlIC} we have computed $\Delta$AIC and $\Delta$BIC with data limited to \texttt{lowl}. Table indicates weak preference for $SA$ when compared with $SA + \beta$ model and a moderate preference for $SA$ when compared with $SA + \alpha + \beta$ model. Lower values of $\Delta$AIC ($\Delta$BIC) of $SA + \beta$ compared to $SA + \alpha + \beta$ can be attributed to it having one less parameter. Significance of models summarized in tables \ref{T:IC} and \ref{T:lowlIC} is the third result of this paper.


\begin{table}
    \begin{tabular}{p{3cm}p{3cm}p{3cm}p{3cm}p{3cm}}
    \hline
    Data & Model & $\chi^2_{\texttt{lowl}}$ & $\Delta$ AIC & $\Delta$ BIC\\
    &&&&\\
    \hline
    \multirow{6}{70pt}{\texttt{TT + lowl + lowE}} & $SA$ & 23.317 & 0 & 0\\
    &&&&\\
    & $SA + \beta$ &  21.467 & 0.150 & 1.481\\
    &&&&\\
    & $SA + \alpha + \beta$ & 22.026 & 2.707 & 5.372 \\
    &&&&\\
    \hline
    \end{tabular}
    \caption{\label{T:lowlIC} The $\Delta$AIC and $\Delta$BIC is computed for \texttt{lowl} data. The table indicates that AIC weakly prefers $SA$ when compared with $SA+\beta$ and moderately prefers $SA$ compared to $SA + \alpha + \beta$. Preference for $SA$ is weak when BIC is used to compare with $SA + \beta$ and moderate when compared with $SA + \alpha + \beta$. }
\end{table}
\begin{table}
 \centering 
 \begin{tabular}{|p{4.52cm}|p{2.7cm}|p{2.7cm}|p{2.7cm}|}
 \hline
 $\texttt{TT + lowl + lowE}$
 & $SA$ & $SA + \beta$ & $SA + \alpha + \beta$  \\
  \hline
 $\Omega_b\,h^2$ & $0.0221\, \pm \, 0.00022$& $0.0221 \pm 0.00022$ & $0.022 \pm 0.00027$ \\
  $\Omega_c\,h^2$ & $0.1206 \pm 0.00210$ & $0.1216 \pm 0.0022$& $0.1221 \pm 0.0022$  \\
 $100\,\theta_{\rm MC}$ & $1.0408 \pm 0.00047$& $1.0407 \pm 0.00047$ & $1.0406 \pm 0.00048$  \\
  $\tau$ & $0.0517 \pm 0.0080$ & $0.0554 \pm 0.0087$ &$ 0.0545 \pm 0.0088$ \\
  ${\rm{ln}}(10^{10} A_s)$ & $3.04 \pm 0.016$ & $3.05 \pm 0.018$ &$3.05 \pm 0.019$  \\
  $n_s$ & $0.9627 \pm 0.0057$ & $0.9573 \pm 0.0067$ &$0.9551 \pm 0.0069$ \\
  $\alpha$ &  0 & 0 & $0.0136 \pm 0.0125$  \\
  $\beta$ &  0 & $0.0133 \pm 0.0089$& $0.0256 \pm 0.0142$\\
 \hline
  $\chi^2$ & $1192.1 \pm 5.5 $ & $1191 \pm 5.6$ & $1190.5 \pm 5.8$\\
  \hline
 $\texttt{CamSpec TT + lowl + lowE}$
 & $SA$ & $SA + \beta$ & $SA + \alpha + \beta$ \\
 \hline
 $\Omega_b\,h^2$ & $0.0221 \pm 0.00020$& $0.0221 \pm 0.00020$& $0.0220 \pm 0.00026$ \\
  $\Omega_c\,h^2$ & $0.1195 \pm 0.00194$ & $0.1202 \pm 0.0020$& $ 0.1205 \pm 0.00208$  \\
 $100\,\theta_{\rm MC}$ & $1.0409 \pm 0.00042$& $1.0408 \pm 0.00042$ & $1.0407 \pm 0.00043$  \\
  $\tau$ & $0.0517 \pm 0.0082$ & $0.0543 \pm 0.0087$ &$ 0.0541 \pm 0.0088$ \\
  ${\rm{ln}}(10^{10} A_s)$ & $3.03 \pm 0.017$ & $3.04 \pm 0.018$&$3.04 \pm 0.018$  \\
  $n_s$ & $0.9630 \pm 0.0056$ & $0.9588 \pm 0.0064$ &$0.9576 \pm 0.0066$ \\
  $\alpha$ &  0 & 0 & $0.0103 \pm 0.0142$  \\
  $\beta$ &  0 & $0.01132 \pm 0.0085$& $0.0208 \pm 0.0155 $  \\
 \hline
  $\chi^2$ & $6848.5 \pm 5.0 $ & $6847.8 \pm 5.1$ & $6848.7 \pm 5.3$\\
  \hline
   $\texttt{HiLLiPoP TT + lowl + }$ $\texttt{LoLLiPoP EE}$  & $SA$ & $SA + \beta $ & $SA + \alpha + \beta$ \\
   \hline
 $\Omega_b\,h^2$ & $0.0222 \pm 0.00019$& $0.0222 \pm 0.00020$& $0.0221 \pm 0.00025$ \\
  $\Omega_c\,h^2$ & $0.1193 \pm 0.00192$ & $0.1197 \pm 0.00197$& $0.1198 \pm 0.00203 $  \\
 $100\,\theta_{\rm MC}$ & $1.0409 \pm 0.00042$& $1.0409 \pm 0.00043$& $1.0408 \pm 0.00044$  \\
  $\tau$ & $0.0572 \pm 0.0062$ & $0.0591 \pm 0.0069$ &$ 0.0590 \pm  0.0069 $ \\
  ${\rm{ln}}(10^{10} A_s)$ & $3.04 \pm 0.014$ & $3.04 \pm 0.016$ &$3.04 \pm 0.016$  \\
  $n_s$ & $0.9642 \pm 0.0053$ & $0.9622 \pm 0.0062$ &$0.9611 \pm 0.0065$ \\
  $\alpha$ &  0 & 0& $0.0045 \pm 0.012$  \\
  $\beta$ &  0 & $0.0057 \pm 0.0085$ & $0.0101 \pm 0.0141$  \\
 \hline
  $\chi^2$ & $11123.0 \pm 6.6  $ & $11123.5 \pm 7$ & $11124.1 \pm 7 $\\
  \hline
 
 \end{tabular}
 \caption{\label{T:pr4_tt} Mean values and standard deviation of cosmological parameters obtained from a Bayesian analysis with different TT and low multipole EE data. We note that while analysis with \texttt{TT + lowl + lowE} data largely agrees with that of \texttt{CamSpec TT + lowl + lowE} (PR4), results from \texttt{HiLLiPoP TT + lowl + LoLLiPoP EE } (PR4) differ.}
 \end{table}
\begin{table}
    \begin{tabular}{p{5cm} p{2cm} p{2cm} p{2cm}}
    \hline
    & $SA$ & $SA + \beta$ & $SA + \alpha + \beta$ \\
  \hline
  \texttt{TT + lowl + lowE} & 34646 & 21064& 17699\\
  &&\\
  \texttt{CamSpec TT + lowl + lowE}& 34924 & 21870 & 19843\\\
  &&\\
  \texttt{HiLLiPoP TT + lowl + LoLLiPoP EE}& 35120 & 28312 & 26374\\
  \hline
    \end{tabular}
    \caption{\label{T:pr4_tt_Shalf} Value of $S_{1/2}$ corresponding to $SA$, $SA + \beta$ and $SA + \alpha + \beta$ models for parameter values given in table \ref{T:pr4_tt}. As in earlier analysis, $SA + \alpha + \beta$ lead to higher amount of suppression. Among different data sets we find that results obtained from analysis of PR3 data using \texttt{Plik} and that of PR4 data using \texttt{CamSpec} are comparable. The analysis of PR4 data using \texttt{HiLLiPoP} and \texttt{LoLLiPoP} likelihoods lead to a higher value of $S_{1/2}$.}
\end{table}
\begin{table}
    \begin{tabular}{p{4cm} p{3cm} p{3cm} p{3cm} p{3cm}}
    \hline
    &  & $SA$ & $SA + \beta$  & $SA + \alpha + \beta$ \\
    \hline 
    \multirow{4}{80pt}{\texttt{TT + lowl + lowE}} & Method I & $1.240\, \pm\, 0.094$& $1.221 \, \pm \, 0.100$  & $ 1.242\,\pm\,0.126$\\
    &&&&\\
    & Method II & - & $1.194\,\pm\,0.091$ & $ 1.142\,\pm\, 0.091$\\
    &&&&\\
    \hline
    \multirow{4}{80pt}{\texttt{CamSpec TT + lowl + lowE}} & Method I & $1.202 \, \pm \, 0.090$& $1.179\,\pm\,0.095$ & $1.196\,\pm\,0.115$\\
    &&&&\\
    & Method II & - & $1.161 \,\pm\, 0.088$ & $1.122\,\pm\, 0.082$\\
    &&&&\\
    \hline
    \multirow{4}{75pt}{\texttt{HiLLiPoP TT + lowl + LoLLiPoP EE}} & Method I & $1.072\, \pm \,0.077$ &  $1.059\, \pm \,0.085$ & $1.060\,\pm\,0.104$\\
    &&&&\\
    & Method II & - & $1.054\,\pm\,0.080$ & $1.036\,\pm\,0.078$\\
    &&&&\\
    \hline
\end{tabular}
\caption{\label{T:pr4_tt_AL} Table of marginalised mean values and the standard deviation of $A_L$ obtained using two methods with PR3 and PR4 datasets and likelihoods. Note that, as found in the previous analysis, mean value of $A_L$ becomes closer to one in $SA + \beta$ and $SA + \alpha+ \beta$  models when Method II is used. This remains true across all datasets and likelihoods. }
\end{table}
\begin{table}
 \centering 
 \begin{tabular}{|p{4.52cm}|p{2.7cm}|p{2.7cm}|p{2.7cm}|}
 \hline
 $\texttt{TTTEEE + lowl + lowE}$
 & $SA$  & $SA + \beta$  & $SA + \alpha + \beta$ \\
  \hline
 $\Omega_b\,h^2$ & $0.0224 \pm 0.00015$& $0.0224 \pm 0.00015$ & $0.0223 \pm 0.00016$ \\
  $\Omega_c\,h^2$ & $0.1202 \pm 0.00136$ & $ 0.1206 \pm 0.00142$& $0.1207 \pm 0.00142$  \\
 $100\,\theta_{\rm MC}$ & $1.0409 \pm 0.00031$& $1.0409 \pm 0.00031$ & $1.0409 \pm 0.00032$  \\
  $\tau$ & $0.0544 \pm 0.0078 $ & $ 0.0573 \pm 0.0085$ &$ 0.0575 \pm 0.0087$ \\
  ${\rm{ln}}(10^{10} A_s)$ & $3.05 \pm 0.016$ & $3.05 \pm 0.018$ &$3.05 \pm 0.018$  \\
  $n_s$ & $0.9649 \pm 0.0043$ & $0.9614 \pm 0.0051 $ &$0.9612 \pm 0.0052$ \\
  $\alpha$ &  0 & 0 & $0.0011 \pm 0.0102$  \\
  $\beta$ &  0 & $0.0102 \pm 0.0084$& $0.0117 \pm 0.0127$\\
 \hline
  $\chi^2$ & $2780.2 \pm 5.8 $ & $2779.8 \pm 5.9$ & $2780.9 \pm 6.1$\\
  \hline
 $\texttt{CamSpec TTTEEE + lowl}$  $\texttt{+ lowE}$
 & $SA$ & $SA + \beta$ & $SA + \alpha + \beta$ \\
 \hline
 $\Omega_b\,h^2$ & $0.0222 \pm 0.00013$& $0.0222 \pm 0.00013$& $0.0222 \pm 0.00014$ \\
  $\Omega_c\,h^2$ & $0.1197 \pm 0.00118$ & $0.1200 \pm 0.00122$& $ 0.1200 \pm 0.00119$  \\
 $100\,\theta_{\rm MC}$ & $1.0408 \pm 0.00025$& $1.0407 \pm 0.00026$ & $1.0408 \pm 0.00025$  \\
  $\tau$ & $0.0517 \pm 0.0079$ & $0.0539 \pm 0.0082$ &$ 0.0541 \pm 0.0083$ \\
  ${\rm{ln}}(10^{10} A_s)$ & $3.04 \pm 0.016$ & $3.04 \pm 0.017$&$3.04 \pm 0.017$  \\
  $n_s$ & $0.9638 \pm 0.0041$ & $0.9607 \pm 0.0049$ &$0.9607 \pm 0.0049$ \\
  $\alpha$ &  0 & 0 & $0.0032 \pm 0.0105$  \\
  $\beta$ &  0 & $0.00891 \pm 0.00809$& $0.0120 \pm 0.0127 $  \\
 \hline
  $\chi^2$ & $10975.9 \pm 5.3  $ & $10975.6 \pm 5.4$ & $10976.6 \pm 5.6$\\
  \hline
   $\texttt{HiLLiPoP TTTEE + lowl +}$ $\texttt{LoLLiPoP EE}$  & $SA$ & $SA + \beta  $ & $SA + \alpha + \beta$ \\
 \hline
 $\Omega_b\,h^2$ & $0.0223 \pm 0.00013$& $0.0223 \pm 0.00013$& $0.0223 \pm 0.00014$ \\
  $\Omega_c\,h^2$ & $0.1189 \pm 0.00118$ & $0.1190 \pm 0.00122$& $0.1189 \pm 0.00122 $  \\
 $100\,\theta_{\rm MC}$ & $1.0409 \pm 0.00026$& $1.0409 \pm 0.00026$& $1.0409 \pm 0.00026$  \\
  $\tau$ & $0.0581 \pm 0.0061$ & $0.0596 \pm 0.0068$ &$ 0.0590 \pm  0.0068 $ \\
  ${\rm{ln}}(10^{10} A_s)$ & $3.04 \pm 0.014$ & $3.04 \pm 0.016$ &$3.04 \pm 0.016$  \\
  $n_s$ & $0.9681 \pm 0.0040$ & $0.9668 \pm 0.0047$ &$0.9669 \pm 0.0048$ \\
  $\alpha$ &  0 & 0& $-0.0068 \pm 0.0095$  \\
  $\beta$ &  0 & $0.00504 \pm 0.00819$ & $-0.0012 \pm 0.0120$  \\
 \hline
  $\chi^2$ & $30574.9 \pm 6.9    $ & $ 30575.3 \pm 7.1$ & $30576.2 \pm 7.3 $\\
  \hline
 \end{tabular}
 \caption{\label{T:pr4_ttteee} Marginalised mean and standard deviation of parameters derived from the posterior probability distribution corresponding to different PR3 and PR4 data sets and likelihoods. Similar to \ref{T:pr4_tt}, results of \texttt{CamSpec TTTEEE + lowl + lowE} are consistent with that of \texttt{TT + lowl + lowE}. Mean value of $\alpha$ and $\beta$ parameters obtained using \texttt{HiLLiPoP + lowl + LoLLiPoP} do not support power suppression.}
 \end{table}
\begin{table}
    \begin{tabular}{p{5cm} p{2cm} p{2cm} p{2cm}}
    \hline
    & $SA$ & $SA + \beta$ & $SA + \alpha + \beta$ \\
  \hline
  \texttt{TTTEEE + lowl + lowE} & 34834 & 23721& 22990\\
  &&\\
  \texttt{CamSpec TTTEEE + lowl + lowE}& 34610 & 24717 & 23576\\\
  &&\\
  \texttt{HiLLiPoP TTTEEE + lowl + LoLLiPoP EE}& 34232 & 28267 & 31145\\
  \hline
    \end{tabular}
    \caption{\label{T:pr4_ttteee_Shalf} Value of $S_{1/2}$ corresponding to marginalised mean values listed in table \ref{T:pr4_ttteee}. We see that analysis of PR4 with \texttt{CamSpec} likelihood is consistent with analysis using PR3 data. Value of $S_{1/2}$ obtained using \texttt{HiLLiPoP + lowl + LoLLiPoP} for $SA + \alpha + \beta$ model is larger than that of $SA + \beta$ model but lower than the  result obtained for $SA$. }
\end{table}
\begin{table}
    \begin{tabular}{p{4cm} p{3cm} p{3cm} p{3cm} p{3cm}}
    \hline
    &  & $SA$ & $SA + \beta$  & $SA + \alpha + \beta$ \\
    \hline 
    \multirow{4}{80pt}{\texttt{TTTEEE + lowl + lowE}} & Method I & $1.183\, \pm\, 0.067$& $1.174 \, \pm \, 0.068$  & $ 1.180\,\pm\,0.070$\\
    &&&&\\
    & Method II & - & $1.165\,\pm\,0.066$ & $ 1.161\,\pm\, 0.066$\\
    &&&&\\
    \hline
    \multirow{4}{85pt}{\texttt{CamSpec TTTEEE + lowl + lowE}} & Method I & $1.092 \, \pm \, 0.060$& $1.082\,\pm\,0.062$ & $1.079\,\pm\,0.061$\\
    &&&&\\
    & Method II & - & $1.078 \,\pm\, 0.058$ & $1.076\,\pm\, 0.059$\\
    &&&&\\
    \hline
    \multirow{4}{95pt}{\texttt{HiLLiPoP TTTEEE + lowl + LoLLiPoP EE}} & Method I & $1.038\, \pm \,0.052$ &  $1.032\, \pm \,0.054$ & $1.035\,\pm\,0.055$\\
    &&&&\\
    & Method II & - & $1.030\,\pm\,0.053$ & $1.036\,\pm\,0.053$\\
    &&&&\\
    \hline
\end{tabular}
\caption{\label{T:pr4_ttteee_AL} Mean and standard deviation of $A_L$ obtained using both methods with various high $\ell$ E mode data.  We find that results obtained using \texttt{CamSpec TTTEEE + lowl + lowE} data is similar to that obtained with \texttt{TTTEEE + lowl + lowE}. Consistent with literature \cite{Tristram:2023haj}, we find that lensing anomaly is significantly reduced in the case of \texttt{HiLLiPoP TTTEEE + lowl + LoLLiPoP EE} data.}
\end{table}
\section{Analysis with Planck Release 4 \label{sec:7}}
Since the release of Planck legacy data a renewed analysis of CMB data, (\texttt{NPIPE}) \cite{Planck:2020olo} taking in to account previously neglected data from the repointing periods and better corrections of systematics,  was released as PR4 data \cite{Tristram:2023haj}. Updated likelihoods have also been proposed to make use of this data. In this section, we shall use PR4 data together with updated likelihoods to investigate power suppression and lensing anomaly. There are primarily three likelihoods, \texttt{CamSpec} \cite{Rosenberg:2022sdy} and High-$\ell$ Likelihood on Polarized Power spectra (\texttt{HiLLiPoP}) for high multipole data \cite{Tristram:2020wbi, Couchot:2015eea} and Low-$\ell$ Likelihood on Polarized Power spectra (\texttt{LoLLiPoP}) for low multipole polarisation data \cite{Tristram:2020wbi, Tristram:2021tvh}. We shall consider four combinations of data namely 
\begin{enumerate}
    \item \texttt{CamSpec TT + lowl + lowE}
    \item \texttt{HiLLiPoP TT + lowl + LoLLiPoP EE}
    \item \texttt{CamSpec TTTEEE + lowl + lowE}
    \item \texttt{HiLLiPoP TTTEEE + lowl + LoLLiPoP EE}
\end{enumerate}

The marginalised mean values and standard deviation of $SA$, $SA + \beta$ and $SA + \alpha +\beta$ for the first two sets of data and corresponding values of $S_{1/2}$ are given in tables \ref{T:pr4_tt} and \ref{T:pr4_tt_Shalf} respectively. The same for latter two data sets are given in tables \ref{T:pr4_ttteee} and \ref{T:pr4_ttteee_Shalf} respectively. Constraints on $A_L$ for these models obtained by analysing first two and last two data sets are listed in tables \ref{T:pr4_tt_AL} and \ref{T:pr4_ttteee_AL} respectively. 
\par 
Our analysis shows that different data sets lead to slightly different constraints on model parameters, leading to different values of suppression. From table \ref{T:pr4_tt_Shalf}, we see that, for all models suppression is largest for \texttt{TT + lowl + lowE}, followed by \texttt{CamSpec TT + lowl + lowE} and \texttt{HiLLiPoP TT + lowl + LoLLiPoP EE}. Suppression obtained with \texttt{CamSpec TT + lowl + lowE} is closer to that obtained with \texttt{TT + lowl + lowE}. We find similar results in table \ref{T:pr4_ttteee_Shalf} for $SA + \beta$ and $SA+\alpha + \beta$ models with an exception that $S_{1/2}$ is lower for $SA + \beta$ than $SA+\alpha + \beta$ while working with \texttt{HiLLiPoP TTTEEE + lowl + LoLLiPoP EE}. 
In the case of $SA$, we find that \texttt{HiLLiPoP TTTEEE + lowl + LoLLiPoP EE} favours parameters with slightly lower power than \texttt{CamSpec TTTEEE + lowl + lowE} which leads to a lower power than \texttt{TTTEEE + lowl + lowE}. 
We note that among parameters obtained from PR3 and PR4 (using \texttt{HiLLiPoP} and \texttt{LoLLiPoP} likelihoods), maximum change is observed in the value of reionization parameter $\tau$.

\par 
As reported in literature, we also find that lensing anomaly is less significant with PR4 data. We find that, for $SA$ when high multipole polarisation data is included, lensing parameter is only $1.5\sigma$ and $0.7\sigma$ away from one while comparing with \texttt{CamSpec TTTEEE + lowl + lowE} and  \texttt{HiLLiPoP TTTEEE + lowl + LoLLiPoP EE} data respectively. This should be compared with $2.7\sigma$ departure observed with \texttt{TTTEEE + lowl + lowE} data. Similarly we find that departure of lensing parameter from one is only $2.24\sigma$ with \texttt{CamSpec TT + lowl + lowE}, $0.93\sigma$ with \texttt{HiLLiPoP TT + lowl + LoLLiPoP EE} compared to $2.5\sigma$ with \texttt{TT + lowl + lowE} data. Note that lensing anomaly is weakest when analysis is conducted with \texttt{HiLLiPoP} and \texttt{LoLLiPoP} likelihoods. Given this observation, we still find that higher the amount of power suppression, more consistent is the value of lensing parameter with one. This is evident from our constraints on $A_L$ obtained using Method II and listed in tables \ref{T:pr4_tt_AL} and \ref{T:pr4_ttteee_AL}. This analysis with PR4 is the fourth result of this paper. 
\section{Forecast for future CMB missions}\label{sec:8}
Bayesian parameter estimation with \texttt{TT + lowl + lowE} constraints $\alpha\, =\, 0.0136 \pm 0.0125$ and $\beta\, =\, 0.0256 \pm 0.0142$ (see table \ref{T:A1}). Though mean value indicates a model which leads to a lack of power at low multipoles, the error bars are quite large with $\alpha\, =\, \beta\, =\, 0$ lying within $2-\sigma$ error bar. Planck satellite has measured temperature anisotropies to great accuracy and it is now only limited by cosmic variance. However this is not the case for polarisation. In this section, we investigate whether a future CMB mission which measures polarisation to a better extent can arrive at tighter constraints on $\alpha$ and $\beta$. For our analysis we shall consider ECHO, a fourth-generation satellite mission proposal for a near-ultimate measurement of the polarisation and discovery of global CMB spectral distortions submitted to Indian Space Research Organisation\footnote{\href{https://www.isro.gov.in}{https://www.isro.gov.in}}. ECHO is similar in spirit to other proposed fourth-generation satellite missions such as COrE \cite{CORE:2017oje}, PRISM \cite{PRISM:2013fvg} and PICO \cite{NASAPICO:2019thw}.
\par 
In order to forecast the extent to which a mission can constrain parameters of a model, we need an estimate of instrument noise. Following \cite{Knox:1995dq, Hobson:1996dg, MAGUEIJ:1997, Magueijo:1996ty, Tegmark:1997vs},
we shall assume a detector with uniform instrumental noise and a Gaussian beam. For such a case, noise power spectrum in a given frequency channel with sensitivity per pixel $\sigma^\nu_{\rm pix}$ and beam with full width at half maximum $\theta^\nu_{\rm FWHM}$ is given by
\begin{equation}\label{eqn:Nl}
  N^{\nu}_{T,\,\ell} = \,\frac{1}{w^\nu \, B_\ell^2(\theta_b)},
\end{equation}
where $1/w^\nu\,=\,(\sigma^\nu_{\rm pix}\times \theta^\nu_{\rm FWHM})^2$ and the Gaussian beam function
\begin{equation}
B_{\ell}(\theta^\nu_b)\, =\, \exp\biggl[-\ell\,(\ell + 1)\,\frac{\theta_b^{\nu^2}}{2}\biggr]
\end{equation}
with $\theta^\nu_b\, =\, \theta^\nu_{\rm FWHM}/\sqrt{8\ln 2}$. 
In writing Eqn. (\ref{eqn:Nl}), we have assumed passive incomplete sky coverage, {\i.e.} a scenario where detector has measured the entire sky, but a fraction of the sky is not considered due to inability to remove foreground signal completely \cite{Magueijo:1996ty}. 
The total noise from all frequency channels is \cite{Valiviita:2017fbx}, 
\begin{equation}
N_{T,\ell}\, = \frac{1}{\Sigma_{\nu} 1/N^\nu_{T,\ell}}.
\end{equation}
 The power spectrum of noise in the polarisation channel is related to that in the temperature channel by $N_{E,\,\ell}\, \propto\, N_{T,\,\ell}$, where in the proportionality constant is two for near-ultimate missions such as ECHO and four for Planck. 

\par
\begin{figure}
    \centering
    \includegraphics[width=0.5\linewidth]{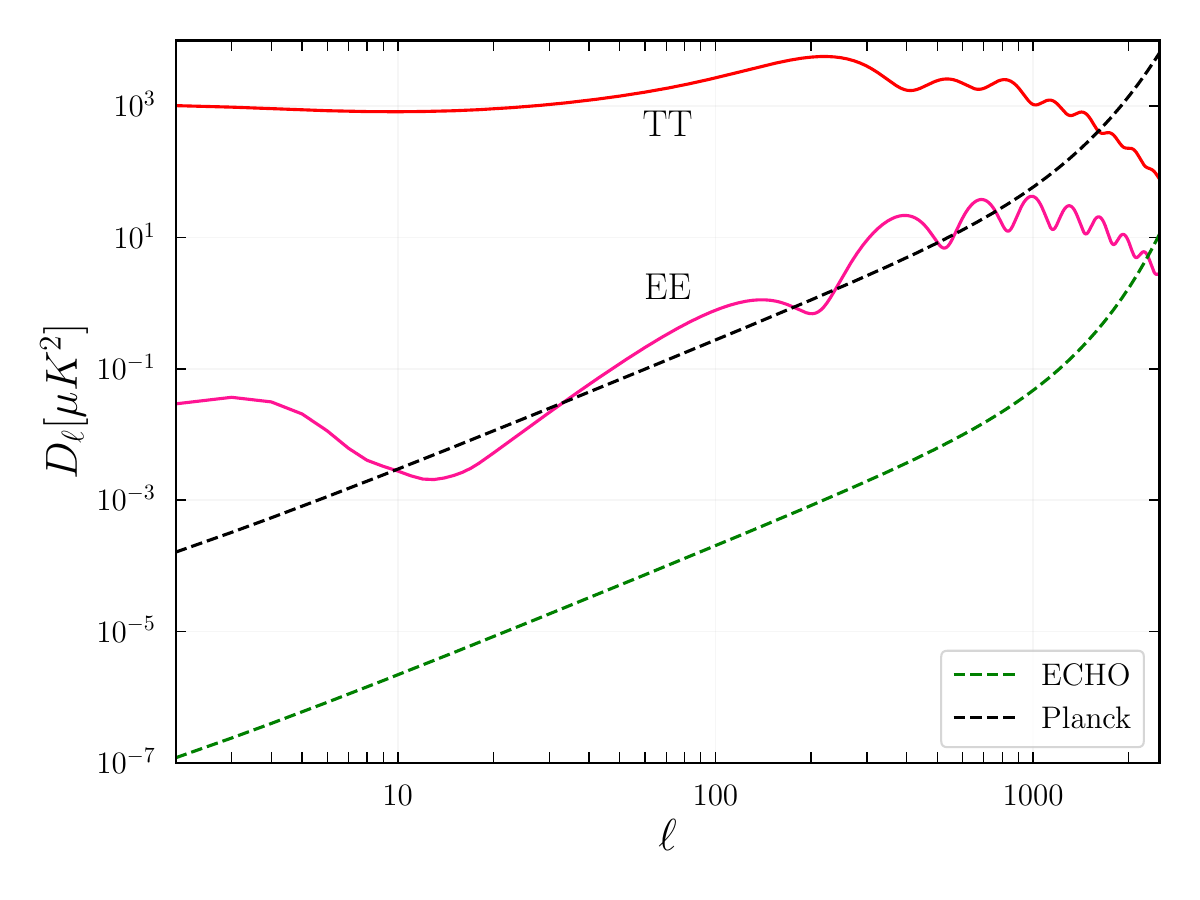}
    \caption{\label{fig:sensitivity} Scale invariant noise power spectra for ECHO and Planck satellites. $D^{TT}_\ell$ and $D^{EE}_\ell$ are given for reference. Note that measurement of E-mode polarisation by ECHO satellite will be near perfect and only be limited by cosmic variance.}
\end{figure}
 The scale invariant spectra of noise in the polarisation channel, $\ell (\ell + 1) N_{E,\ell}/2/\pi $ for ECHO and Planck missions are given in figure \ref{fig:sensitivity}. For this, we have used the values of $1/w^\nu$ provided in the proposal by CMB-Bh$\overline{\rm a}$rat consortium (also given in table 1 of \cite{Adak:2021lbu}). ECHO has a total of $20$ frequency channels for measuring polarisation. We consider five frequency channels in the range 100-200 GHz which we consider as relevant for extracting cosmological information. For obtaining noise spectra of Planck, we have worked with values of $\sigma_{\rm pix}$ and $\theta_{\rm FWHM}$  provided as $\Delta T/ T\, (\mu K/K)$ and angular resolution respectively, in the Planck blue book (table 1.1 of \cite{Planck:2006aa}).  Out of the nine frequency channels, we focus on channels with frequency $100$ GHz, $143$ GHz and $217$ GHz that are relevant for cosmology.
We have also plotted $D^{TT}_{\ell}$ and $D^{EE}_{\ell}$ for reference. In obtaining these curves, we have assumed $SA$ model with parameter values set to marginalised mean values obtained up on comparison with \texttt{TT + lowl + lowE} as the fiducial model (see table \ref{T:A1}). 
\par 
\begin{figure}
         \centering
        \begin{tabular}{cc}
        \includegraphics[width=0.6\linewidth]{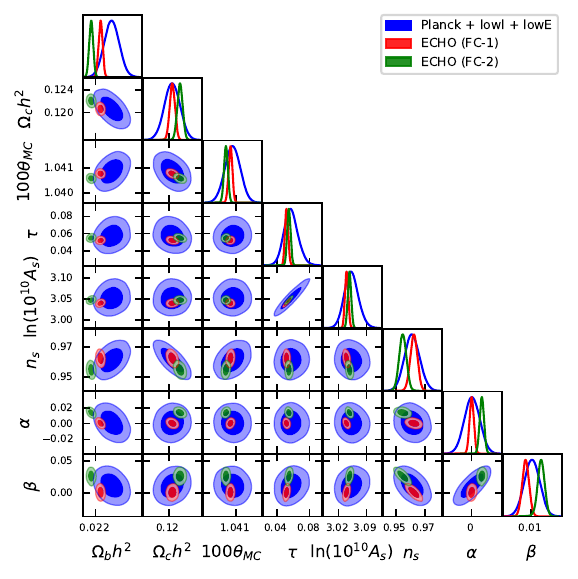}&
        \includegraphics[width=0.36\linewidth]{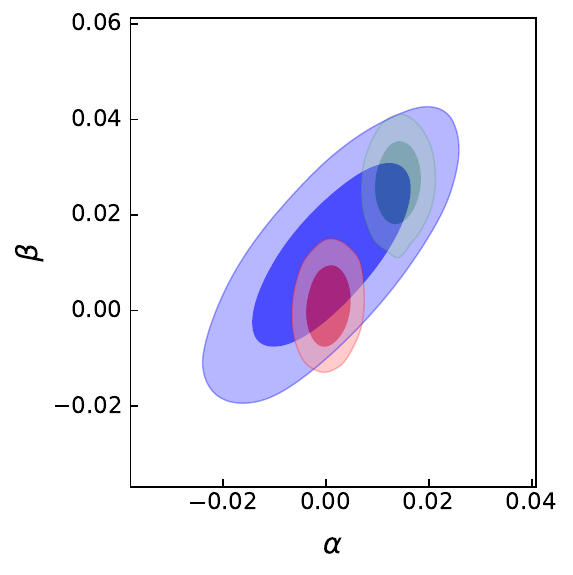}
        \end{tabular}
        \caption{\label{fig:forecast}
        Forecast for model parameters of $SA + \alpha + \beta$ model that can be obtained with ECHO mission is shown in the left. We have considered two fiducial cosmologies without (FC-1) and with (FC-2) power suppression at low multipoles. Forecast of joint posterior distribution of $\alpha$ and $\beta$ has been magnified and plotted on the right. From right panel it is very much clear that the presence of power suppression can be estimated much more decisively from near-ultimate CMB missions such as ECHO.}
\end{figure}
To forecast achievable constraints on $\alpha$ and $\beta$ with ECHO mission, along with the above noise model, we work with two fiducial cosmologies. Firstly, we consider angular power spectra generated with $SA$ model with parameter values set to their marginalised mean values obtained up on comparison with \texttt{TT + lowl + lowE}, see table \ref{T:A1}. We refer to this data as Fiducial Cosmology I (FC-1). This model implies no suppression of power at low multipoles. Secondly, we consider angular power spectra generated by $SA + \alpha + \beta$ model with parameter values set to marginalised mean values obtained up on comparison with \texttt{TT + lowl + lowE}, see table \ref{T:A1}. We refer to it as Fiducial Cosmology II (FC-2). 
Forecasts for $\alpha$ and $\beta$ are obtained using Bayesian parameter estimation with data described by the fiducial cosmology and the model of noise. We have used exact likelihood provided by \verb|CosmoMC| for this analysis, see, for instance, \cite{Hamimeche:2008ai}. This likelihood incorporates cosmic variance and we have assumed a usable sky fraction, $f_{sky}\,=\,0.7$. The parameter forecasts for $SA + \alpha + \beta$ model for both fiducial cosmologies are given in the corner plot \ref{fig:forecast}. We have included the constraints obtained from \texttt{TTTEEE + lowl + lowE} for comparison. An enlarged view of joint confidence contours of $\alpha$ and $\beta$ parameter is also given. We find that confidence contours of cosmological parameters are smaller for ECHO mission which illustrates its ability to arrive at tighter constraints. We find that, for FC-2 fiducial cosmology, forecasted values of $\alpha$ and $\beta$ are $\alpha\, =\, 1.400\times 10^{-2} \pm 2.873 \times 10^{-3}$ and $\beta\, =\, 2.676\times 10^{-2} \pm 5.587 \times 10^{-3}$. Such a measurement can rule out null values for $\alpha$ and $\beta$  by more than $4-\sigma$. 
Further, it is evident from figure \ref{fig:forecast} that, when working with FC-1 the values of 
$\alpha$ and $\beta$ are centred around zero. 
This establishes ability of future CMB missions such as ECHO to make stronger conclusions about the presence of power suppression using simple phenomenological models that describe spectra at low multipoles. This is the final result of this paper. 
\section{Summary and Discussion}\label{sec:9}
In this work, we investigated evidence for lack of power at low multipoles and its connection to lensing anomaly. We considered three simple extensions to standard ansatz ($SA$) involving running and running of running of spectral tilt, namely $SA + \alpha$, $SA + \beta$ and $SA + \alpha + \beta$ and compared predictions of these models with different data sets including Planck legacy data, BAO, DES and also the latest PR4 with updated likelihoods namely \texttt{CamSpec}, \texttt{HiLLiPoP} and \texttt{LoLLiPoP}. It should be highlighted that these extensions to $SA$ are agnostic about power suppression at low multipoles.
\par 
We performed four different analyses. First, we performed a Bayesian parameter estimation of these models with CMB temperature, polarisation, BAO and DES data sets.  This analysis was described in sections  \ref{sec:4} and \ref{sec:7}. 
This analysis showed that marginalised mean values of parameters of extended models, derived from their posterior distribution leads to a suppression of power at low multipoles compared to that obtained in $SA$. 
This is evident from our computations of $S_{1/2}$ given mainly in tables \ref{T:Shalf}, \ref{T:pr4_tt_Shalf} and \ref{T:pr4_ttteee_Shalf}. From these estimations  performed with different data sets, we also find that the suppression is most in $SA + \alpha + \beta$, followed by $SA + \beta$. The only exception to this is in the case of analysis with \texttt{HiLLiPoP TTTEEE + lowl + LoLLiPoP EE} where $SA + \beta$ model leads to the lowest value of $S_{1/2}$. 
Our analysis with \texttt{TT + lowl + lowE} data indicates that among the extended models, $SA+\alpha$ has the most power at low multipoles. We also find that, barring minor exceptions as explained in section \ref{sec:4}, addition of high-$\ell$ polarisation, BAO or DES data leads to parameters whose mean values lead to a higher value of $S_{1/2}$. 
\par 
In the second analysis, performed in sections \ref{sec:5} and \ref{sec:7}, we investigated alleviation of lensing anomaly in extensions to $SA$. 
We did this following two methods. In Method I, we did a Bayesian parameter estimation of extensions to $SA$ after including $A_L$ as a free parameter. In Method II, we fixed the parameters $\alpha$ and $\beta$ to their marginalised mean values and varied the resultant seven parameter model. 
Our analysis indicates that lensing anomaly is alleviated in $SA+\beta$ and $SA + \alpha + \beta$ models. In Method I, where all parameters including $\alpha$ and $\beta$ are varied, the reduction of lensing anomaly occurs through a lowering of mean value of $A_L$ or a broadening of error bars. 
When Method II is used we find that mean value of $A_L$ decreases, thus alleviating the lensing anomaly. \newvchanges{ This alleviation is due to  the negative correlation of the suppression parameters 
$\alpha$ and $\beta$ with the lensing parameter $A_L$. As a consequence of this, positive values of $\alpha$ and $\beta$, which induce a suppression of power, lead to lower values of $A_L$.} We find that the higher the amount of power suppression more the mitigation of tension in lensing parameter. 
\par
\begin{figure}
         \centering
        \begin{tabular}{cc}
        \includegraphics[width=0.45\linewidth]{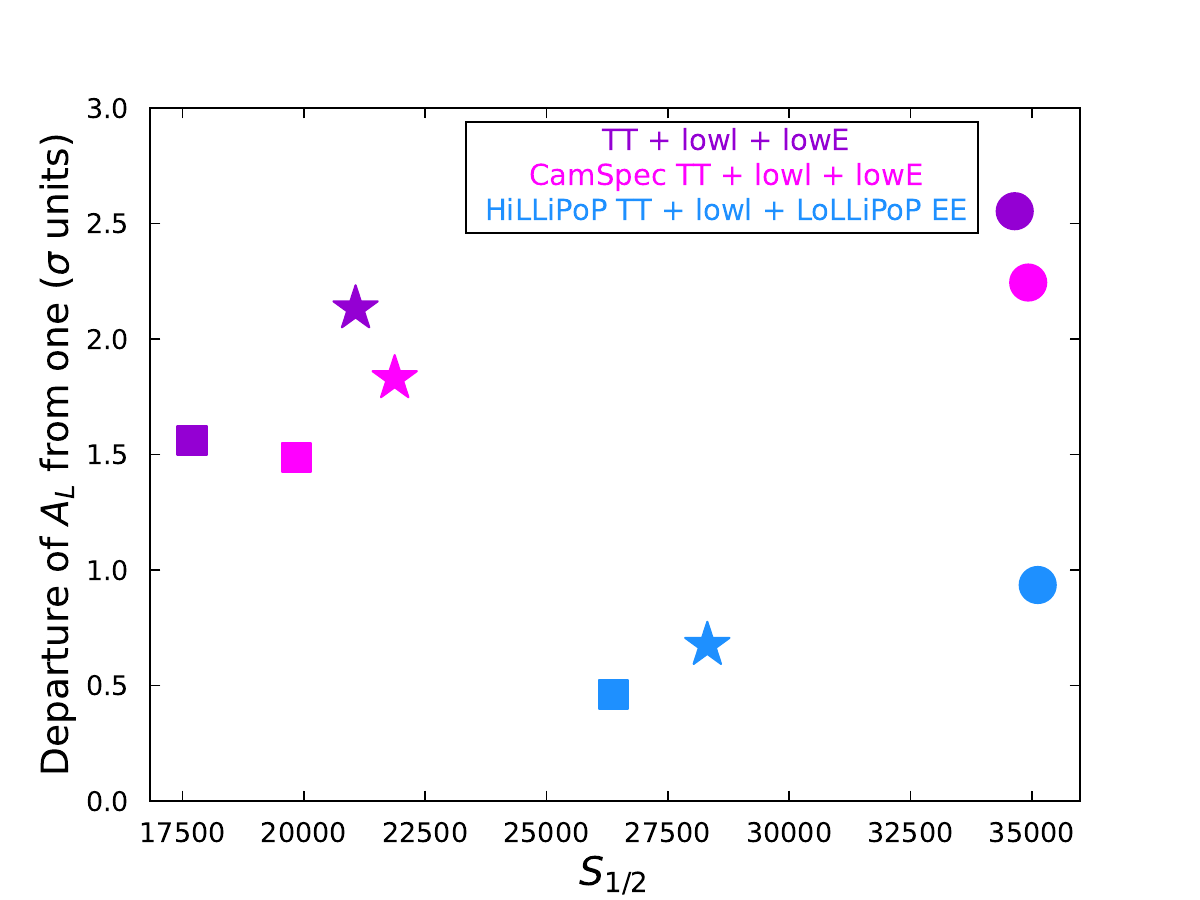}&
        \includegraphics[width=0.45\linewidth]{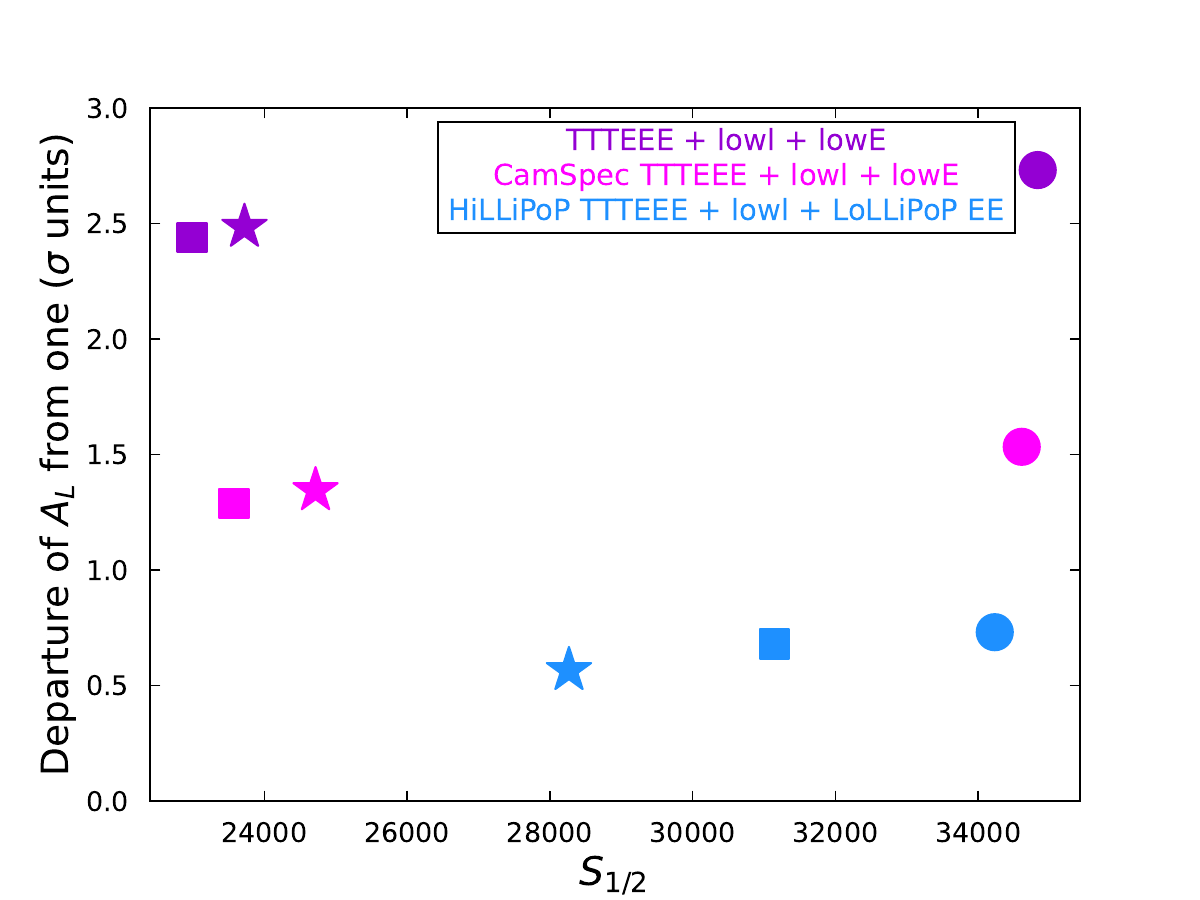}
        \end{tabular}
        \caption{\label{fig:summary} Plot of departure of $A_L$ from one in terms of standard deviation ($\sigma$) versus $S_{1/2}$ without (left) and with high $\ell$ polarisation (right). Circles, stars and squares represent $SA$, $SA + \beta$ and $SA + \alpha + \beta$ models respectively. $S_{1/2}$ values are drawn from tables \ref{T:pr4_tt_Shalf} and \ref{T:pr4_ttteee_Shalf}. $A_L$ values are drawn from tables \ref{T:pr4_tt_AL} and \ref{T:pr4_ttteee_AL}. We see that for any given data set, lower the power at low multipoles, lower the departure of $A_L$ from one.}
\end{figure}

Results of the above two analyses obtained using different releases of Planck data and likelihood codes are summarised in figure \ref{fig:summary}. In the figure, $x-$axis contains the value of $S_{1/2}$ obtained using the first analysis, and the $y-$axis contains departure of lensing parameter from one in terms of standard deviation obtained using Method II described above. From the figure, we see that with PR4 data lensing anomaly is significantly reduced, even in the case of $SA$. This is seen more so with \texttt{HiLLiPoP TTTEEE + lowl + LoLLiPoP EE} data \cite{Tristram:2023haj}. We also find that compared to PR3 data, PR4 selects marginalised mean value of parameters of extended models which lead to higher value of $S_{1/2}$. Given this reduction in lensing anomaly with PR4 data, we still find that value of $S_{1/2}$ and $A_L$ are positively correlated.  We note that this correlation is strongest when high multipole polarisation data is not included. 
\par 
Thirdly we approached the problem from the perspective of information criteria. We find that Akaike information criteria weakly prefers $SA$ over $SA + \beta$ and $SA + \alpha + \beta$ models, see table \ref{T:IC}. The preference slightly increases for $SA$ against $SA + \alpha + \beta$ when polarisation data is used. Addition of BAO or DES do not affect the analysis. 
Bayesian information criteria penalises addition of extra parameters. This results in $SA$ being strongly preferred against both $SA+\beta$ and even more so against  $SA + \alpha + \beta$ models.  However, if we limit the analysis of BIC to \texttt{lowl} data, the preference for $SA$ against $SA+\beta$ becomes weak and that against $SA + \alpha + \beta$ becomes moderate, see table \ref{T:lowlIC}. We believe that it is important to limit data to relevant multipoles when investigating relevance of a local feature such as power suppression. 
In this analysis, we have not explicitly evaluated $\Delta$AIC and $\Delta$BIC with  PR4 as we expect the results to be largely the same as those with PR3 data. 
\par 
From figure \ref{fig:summary}, we see that high-$\ell$ polarisation data prefers parameters with lesser amount of power suppression. Though PR4 data have tried to improve the analysis, it should be underlined that measurement of polarisation by Planck is noisy, see figure \ref{fig:sensitivity}. Hence, in the fourth analysis, we predict the constraints on model parameters that can be obtained by near-ultimate future CMB space mission such as ECHO. Since the noise in measuring polarisation by ECHO is much smaller than the signal, it can constrain parameters to a greater extent than Planck. Figure \ref{fig:forecast} provides our prediction of the posterior probability distribution of model parameters that can be obtained by ECHO. We considered two fiducial cosmologies, namely one without (FC-1) and another with  power suppression (FC-2). We assumed a uniform white noise for our analysis. We find that if we assume a fiducial cosmology with power suppression, it is able to constrain $\alpha$ and $\beta$ with null values excluded at more than $4-\sigma$. This illustrates the ability of future CMB missions to ascertain the presence or absence of a power suppression at low multipoles with a simple $SA + \alpha + \beta$ model. 
\par  
The conclusions of our analysis can be stated as follows. We find that there is some evidence for the existence of power suppression at low multipoles. We find that power suppression in these models lead to alleviation of lensing anomaly. 
The strength of the evidence for power suppression varies with data sets and likelihood codes. We believe that evidence for a local feature should be computed using data around that feature. Finally, and most importantly, we can ascertain the existence or absence of power suppression with future CMB missions. 
\par 
Before we conclude it is interesting to relate our findings to existing literature, mention certain caveats and discuss scope for future exploration. It has been pointed out that lensing anomaly gets alleviated if we consider very high multipole data such as from ACT \cite{ACT:2020gnv}. 
However, this does not explain why Planck data is inconsistent with $A_L\,=\,1$. In this context, it is also important to note that 
main source of lensing anomaly is from data at low-$\ell$ \cite{Ben-Dayan:2024uvx}. In our analysis, we notice that the information criteria penalises the number of parameters in a given model. Hence, if a model derives its suppression of power from some fundamental physics such that it does not require extra parameters, then  such a model can be more preferred. Models, we have considered here, are agnostic about power suppression. The priors we have considered in table \ref{T:prior} can equally generate a power spectrum with enhancement as well as suppression of power at low multipoles. It is true that, for these models, power suppression at low multipoles implies an enhancement in power at higher multipoles. However, when limited to low multipoles, say $\ell \lesssim 2500$, the extended models mainly differ with $SA$ in their behaviour at low multipoles. Though we have used $SA + \beta$ and $SA + \alpha + \beta$ for this study, we do not advocate that this is the underlying model of early universe but have only used this as a convenient template. As indicated in figure \ref{fig:ClCtheta}, these models lead to a gradual suppression of power. It would be interesting to extend this study to models with more nuanced features. Different datasets involve different nuisance parameters. It would also be interesting to study their correlation with other cosmological parameters. Finally and more importantly, this study establishes the need for advanced CMB missions such as ECHO and the ability of simple models such as $SA + \alpha + \beta$ to test the presence of power suppression with such missions. 

\section*{Acknowledgement}
We thank L. Sriramkumar for his valuable comments. We acknowledge Anusandhan National Research Foundation (ANRF) for their support through Start-up Research Grant SRG/2021/001769. We acknowledge the use of PU HPC facility of the National Supercomputing Mission and thank Centre for Cyber Physical Systems, National Institute of Technology Karnataka, Surathkal for the financial support for the same.

\bibliographystyle{unsrt}


\end{document}